\documentclass[aps,prx,reprint,superscriptaddress]{revtex4-2}
\usepackage{graphicx}
\usepackage[dvipsnames]{xcolor}
\usepackage[colorlinks=true,linkcolor=Blue,citecolor=Blue,urlcolor=Blue,linktocpage=true]{hyperref}
\usepackage{amsmath,amssymb,amscd,amsfonts,bm, bbm}
\usepackage[capitalize]{cleveref}
\usepackage{parskip}
\usepackage{siunitx}
\usepackage{mleftright}
\mleftright

\usepackage{mathtools}
\DeclarePairedDelimiter{\ket}{\lvert}{\rangle}%
\DeclarePairedDelimiterX\innerp[2]{\langle}{\rangle}{#1\delimsize\vert\mathopen{}#2}%
\DeclarePairedDelimiterX\braket[2]{\langle}{\rangle}{#1\delimsize\vert\mathopen{}#2}%
\DeclarePairedDelimiterX\braketOP[3]{\langle}{\rangle}{#1\,\delimsize\vert\,\mathopen{}#2\,\delimsize\vert\,\mathopen{}#3}%
\DeclarePairedDelimiterX\ketbra[2]{\lvert}{\rvert}{#1\delimsize\rangle\!\delimsize\langle#2}%
\DeclarePairedDelimiterX\outerp[2]{\lvert}{\rvert}{#1\delimsize\rangle\!\delimsize\langle#2}%
\DeclarePairedDelimiterX\projector[1]{\lvert}{\rvert}{#1\delimsize\rangle\!\delimsize\langle#1}%
\DeclarePairedDelimiterX\dyad[2]{\lvert}{\rvert}{#1\delimsize\rangle\!\delimsize\langle#2}%
\DeclarePairedDelimiterX\comm[2]{[}{]}{#1,#2}
\newcommand{\mel}[3]{\braketOP{#1}{#2}{#3}}

\newcommand*{\scrpt}[1]{\mathrm{#1}}
\newcommand*{\s}[1]{\ensuremath{_\scrpt{#1}}}

\begin{document}

	\newcommand\thetitle{Heralded quantum non-Gaussian states in pulsed levitating optomechanics}
	\title{\thetitle}

	\author{F. Bemani}
	\email{foroudbemani@gmail.com}
	\affiliation{Department of Optics, Palack\'{y} University, 17. listopadu 1192/12, 77146 Olomouc, Czech Republic}

	\author{A. A. Rakhubovsky}
	\email{rakhubovsky@optics.upol.cz}
	\affiliation{Department of Optics, Palack\'{y} University, 17. listopadu 1192/12, 77146 Olomouc, Czech Republic}

	\author{R. Filip}
	\email{filip@optics.upol.cz}
	\affiliation{Department of Optics, Palack\'{y} University, 17. listopadu 1192/12, 77146 Olomouc, Czech Republic}

	\date{\today}

	\begin{abstract}
		Optomechanics with levitated nanoparticles is a promising way to combine very different types of quantum non-Gaussian aspects induced by continuous dynamics in a nonlinear or time-varying potential with the ones coming from discrete quantum elements in dynamics or measurement.
		First, it is necessary to prepare quantum non-Gaussian states using both methods.
		The nonlinear and time-varying potentials have been widely analyzed for this purpose.
		However, feasible preparation of provably quantum non-Gaussian states in a single mechanical mode using discrete photon detection has not been proposed yet for optical levitation.
		We explore pulsed optomechanical interactions combined with non-linear photon detection techniques to approach mechanical Fock states and confirm their quantum non-Gaussianity.
		We also predict the conditions under which the optomechanical interaction can induce multiple-phonon addition processes, which are relevant for $n$-phonon quantum non-Gaussianity.
		The practical applicability of quantum non-Gaussian states for sensing phase-randomized displacements is shown.
		Besides such applications, generating quantum non-Gaussian states of levitated nanoparticles can help to study fundamental questions of quantum thermodynamics, and macroscopic quantum effects.
	\end{abstract}


	\maketitle

	\section{\label{Sec:Sec1} Introduction}
	Quantum non-Gaussian (QNG) states of linear harmonic oscillators are essential for quantum science and technology~\cite{walschaers_nongaussian_2021}.
  They are already known as a prerequisite for universal quantum simulation~\cite{shen_quantum_2018,wang_efficient_2020,dutta_simulating_2024} and computation~\cite{menicucci_universal_2006,mari_positive_2012}, quantum error correction~\cite{michael_new_2016,cai_bosonic_2021} and quantum metrology advantage~\cite{strobel_fisher_2014,deng_quantumenhanced_2024}.
	The truly quantum non-Gaussianity originating from genuinely quantum forms of nonlinear operations has to be distinguished from incoherent statistical mixtures of Gaussian states provided by nonlinearities in semiclassical regimes.
	Subsequently, QNG phenomena are crucial for many fundamental quantum information protocols since they are a resource for wide applications in quantum sensing~\cite{tan_nonclassical_2019} and advancement of quantum information processing~\cite{gan_hybrid_2019,gao_entanglement_2019}.
As a useful tool for the pioneering experiments, it has been proposed to exploit the Fock basis statistics for evaluation of the QNG states~\cite{filip_detecting_2011,lachman_quantum_2022}.
	The main idea is that Gaussian states have limits to be restricted in a finite Fock-basis subspace.
	This also applies to states made by Gaussian operations on Fock states, creating a hierarchy of multiphoton quantum non-Gaussianity~\cite{podhora_quantum_2022}.
	The methods of evaluation of quantum non-Gaussianity have been verified in various optical, atomic and solid-state experiments~\cite{jezek_experimental_2011,jezek_experimental_2012,straka_quantum_2014,lachman_faithful_2019,higginbottom_pure_2016,mika_singlemode_2022,predojevic_efficiency_2014,straka_quantum_2018,podhora_quantum_2022} to detect QNG states with positive Wigner functions affected by noise and loss.

	Micro- or nano-sized particles with mass in the range of femtograms to picograms trapped in optical tweezers--tightly focused laser beams that can trap and manipulate particles through radiation pressure and gradient forces-- form a promising platform for investigations of light-matter interaction~\cite{millen_optomechanics_2020,gonzalez-ballestero_levitodynamics_2021,winstone_levitated_2023}.
	The dipole force experienced by the particle is proportional to the gradient of the trapping field intensity.
	Therefore, the spatial intensity profile of the tweezer serves as an effective variable potential for mechanical motion, providing a harmonic trap with a frequency in the range between tens and hundreds of~\unit{\kilo\hertz}.
	No mechanical contact with the environment, excellent control over the frequency and position of the trapping laser, and free rotation of the levitated particles have opened the way to impressive experimental progress and applications~\cite{millen_cavity_2015}.
	Studying macroscopic quantum mechanics requires cooling trapped NPs close to their ground states~\cite{millen_cavity_2015,romero-isart_optically_2011,chang_cavity_2010} which is possible using techniques of cavity cooling~\cite{windey_cavitybased_2019,delic_cavity_2019,delic_cooling_2020,gonzalez-ballestero_theory_2019} or feedback~\cite{gieseler_subkelvin_2012,li_millikelvin_2011,tebbenjohanns_motional_2020}.
	These trapped NPs are suitable platforms for ultrasensitive detection of displacement, force, acceleration, and torque~\cite{millen_optomechanics_2020,weiss_large_2021,geraci_sensing_2015,ranjit_zeptonewton_2016,monteiro_optical_2017,hebestreit_sensing_2018,hoang_torsional_2016,ahn_ultrasensitive_2020,iakovleva_zeptometer_2023}, and fundamental physics tests~\cite{romero-isart_large_2011,rudolph_entangling_2020,chauhan_stationary_2020,rakhubovsky_detecting_2020,carney_testing_2021,carney_searches_2023}.

  Here, we propose a protocol to create QNG Fock states of mechanical center-of-mass motion of a levitated nanoparticle (NP) in the experimentally feasible regime using discrete nonlinear detectors.
  While the general approach of single-phonon addition/subtraction has been established in prior studies, its direct application to levitodynamics with its peculiarities with a view toward quantum non-Gaussianity has remained elusive.
  Our work aims to bridge this gap.
  Specifically, we contribute to the theoretical and practical understanding of QNG states in levitated systems through several novel aspects: We develop a theoretical framework to analyze QNG state preparation in the pulsed regime and extend the model beyond the rotating wave approximation.
  We adapt existing methods to generalize quantum non-Gaussianity to multi-phonon states, addressing unique challenges in optical levitation.
  We extend the theoretical framework to enable QNG state generation in cavityless levitated systems, a previously unexplored regime.
  Finally, we propose a position sensor leveraging the non-Gaussian properties of mechanical states, demonstrating their practical applicability for sensing phase-randomized displacements --- an application not previously discussed.
  Collectively, these contributions bridge the gap between established discrete quantum control methods and the tailored requirements of levitated optomechanics, marking a significant advancement in both theoretical and experimental domains.
  The generated motional mechanical states can be conclusively verified using specific quantum non-Gaussianity criteria of Refs.~\cite{lachman_faithful_2019,innocenti_nonclassicality_2022}.
  Optical readout of the mechanical states ~\cite{vanner_optomechanical_2015}, combined with a loss-tolerant version of the QNG criteria ~\cite{lachman_faithful_2019,innocenti_nonclassicality_2022}, facilitates direct observation of QNG features using photon-counting detectors, which differs from linear detection methods like homodyne or heterodyne measurement.
  Our protocol circumvents the need for retrodiction of mechanical QNG states from optical states, allowing for straightforward and reliable verification of QNG features.
  Our work establishes three pivotal milestones.
  First, we demonstrate the feasibility of obtaining a QNG state of the mechanical oscillator sufficiently robust against environmental noise to allow its subsequent optical readout despite experimental imperfections, such as initial state impurity and heating.
  Second, we verify that this non-Gaussianity can be observed directly in the optical output of the optomechanical system, avoiding indirect inference methods.
  Finally, we confirm the potential use of these states for a major application of levitating particles --- quantum sensing of external forces.

	\section{Results}
	\begin{figure}
		\includegraphics[width=\columnwidth]{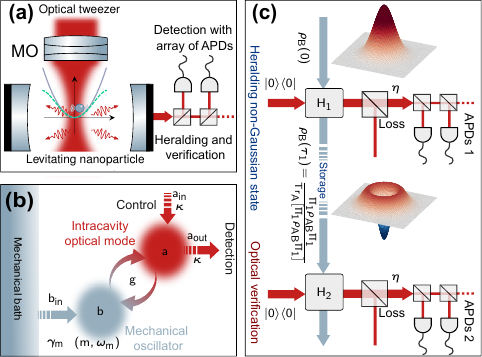}
		\caption{
			(a) Illustration of the proposed heralded generation of QNG states for an optomechanical system with a levitated NP trapped within an optical cavity by an optical tweezer.
			The motion of the NP is coupled to the cavity field via coherent scattering.
			(b) Experimental implementation of single-phonon addition or subtraction with known experimental parameters and technical imperfections.
			(c) The protocol and theory model for generating and verifying QNG mechanical states in a pulsed optomechanical system.
			The system comprises a cooled mechanical oscillator (denoted by annihilation operator $b$) coupled to an optical cavity mode (denoted by $a$).
			The coupling is enabled by pulsed laser light and lasts for a duration $\tau_1$ with Hamiltonian $H_1$ (beam splitter or parametric squeezing interaction).
			Upon a photon detection at the array of APDs 1, the mechanical state is projected onto a non-Gaussian quantum state.
			The generated mechanical state is then transferred to the optical mode via the beam splitter interaction $H_2$ for a duration $\tau_2$ and verified at the array of APDs 2.
		}
		\label{Fig1}
	\end{figure}

	\subsection{Levitated Nanoparticle Protocol}
	As depicted in Fig.~\ref{Fig1}~(a), we propose a scheme to generate and verify QNG mechanical states in a pulsed optomechanical system.
	The system consists of a levitated NP coupled to an optical cavity.
	The coupling could be realized via radiation pressure~\cite{kiesel_cavity_2013}, however, here we focus on the coherent-scattering-based approach~\cite{windey_cavitybased_2019}.
	The interaction between the motion of the NP and the electromagnetic fields is described by a Hamiltonian $ H_{\rm int} = -\frac{1}{2}\alpha_p {\bf E}^2(x_m,t)$ that depends on the polarizability of the NP $\alpha_p$ and the electric field ${\bf E}(x_m,t)$ at the center of mass position of the particle $x_m$~\cite{gonzalez-ballestero_theory_2019}.
	This electric field can be decomposed into contributions from the tweezer field and the cavity field, ${\bf E} = {\bf E}_{\rm tw} + {\bf E}_{\rm cav}$.
	When this decomposition is substituted back into the interaction Hamiltonian, three terms emerge.
	Two of these terms, ${\bf E}_{\rm tw}^2$ and ${\bf E}_{\rm cav}^2$, are proportional to the square of the tweezer field and the square of the cavity field, respectively.
	The former gives rise to the trapping potential, while the latter corresponds to the radiation pressure optomechanical interaction, which can be neglected or avoided by carefully placing the NP at the node of the standing wave within the cavity.
	The cross term ${\bf E}_{\rm tw}\cdot{\bf E}_{\rm cav}$ describes the coherent scattering of tweezer photons into the cavity field.
	The cavity field can be expressed as ${\bf E}_{\rm cav}={\bf E_c}( a+ a^\dagger)$ in terms of the field magnitude ${\bf E_c}$ that depends on the volume of the cavity mode and the normalized classical electric field of the cavity mode, and the creation ($a^\dagger$) and annihilation ($a$) operators.
	The total Hamiltonian with $\hbar=1$ is then given by~\cite{gonzalez-ballestero_theory_2019}
	\begin{equation}\label{Eq:Lin_Hamiltonian}
		H = \Delta a^\dagger a + \frac{p_m^2}{2m} + \frac{1}{2}m\omega_m^2x_m^2  - G(a+a^\dagger)x_m,
	\end{equation}
	In this expression $\Delta = \omega_{\rm cav} - \omega_{\rm tw}$ is the detuning between the cavity and tweezer frequencies, $p_m$ the NP momentum operator, and $m$ its mass.
	The specific values of the trapping frequency $\omega_m$ and the coupling rate $G$ in the Hamiltonian depend on the properties of the NP, the tweezer field, and the cavity mode.
	These properties must be carefully chosen to realize the desired QNG mechanical states in the optomechanical system.
	In principle, alternative platforms, such as particles levitated in radio-frequency Paul traps~\cite{goldwater_levitated_2019,bonvin_hybrid_2024,dania_ultrahigh_2024} or magnetically levitated NPs~\cite{hofer_highq_2023} can be used, however, here we focus on the optically levitated NPs as the platform with the most advanced cooling of the center of mass motion reported to date.

	Our approach to generating non-Gaussian states involves employing non-Gaussian measurements on the optical field, which is correlated with the mechanical mode.
	Specifically, photon-counting measurements facilitated by APDs or equivalent detectors can project the mechanical state onto a QNG mechanical state, albeit conditionally.
	In fact, at the detector, a measurable charge pulse is obtained when a photon hits it.
	In a linear detector, the resulting current is directly proportional to the incoming photon flux.
	However, when detectors operate at very low intensities, they use an avalanche process to transform a single ionization event into a detectable pulse.
	Therefore, it is only possible to distinguish between the vacuum state and the occupied states.
	The action of these binary on-off detectors is represented by a two-value positive operator-valued measure (POVM)~\cite{serafini_quantum_2017}
	\begin{equation}
		{\Pi _0}(\eta ) = \sum\limits_{n = 0}^\infty  {{{(1 - \eta )}^{n}}\left| n \right\rangle \left\langle n \right|}, \quad {\Pi _1}(\eta ) = 1 - {\Pi _0}(\eta ),
	\end{equation}
	where $\eta$ is the detection efficiency.
  We consider the electronic noise and dark counts to be negligible for the current detectors.
  An advantage of the coherent-scattering-based setup for our purpose is that only the inelastically scattered photons populate the cavity mode and can eventually reach the detector.
  Stray tweezer photons do not get scattered to the cavity and are therefore automatically rejected.
  This is in contrast with other optomechanical systems, where drive photons have to be carefully filtered out using convoluted optical filters~\cite{enzian_nongaussian_2021,patil_measuring_2022a}.
	In the event of a click, the optomechanical state $\rho_{\text{AB}}$ is projected onto the conditional state of the mechanics:
	\begin{equation}
		\rho_{\text{B}}(\tau_1) = \frac{\Pi_1 \rho_{\text{AB}} (\tau_1)\Pi_1}{\text{Tr}_A[\Pi_1 \rho_{\text{AB}}(\tau_1)\Pi_1]}\,,
	\end{equation}
	where $\text{Tr}_A(\cdot)$ denotes the trace over the optical mode.

	Phonon subtraction/addition protocols involve a weak, pulsed optomechanical interaction followed by photon detection using an APD.
	For phonon subtraction, a beamsplitter-like interaction
	\begin{equation}\label{Eq:BSHamiltonian}
		{H_{{\rm{BS}}}} = -g(a{b^\dag } + b{a^\dag })\,,
	\end{equation}
	with vacuum in the optical input is used, realizable by driving the system on the lower mechanical sideband ($\Delta=\omega_m$)~\cite{aspelmeyer_cavity_2014}.
	Here we have introduced the mechanical annihilation operator $b = \sqrt {m{\omega _m}/2} \left[ {{x_m} + i{p_m}/(m{\omega _m})} \right]$ and $g = G/\sqrt{2m\omega_m}$.
	Assuming unitary dynamics and weak coupling $g \tau_1 \ll 1$, the transformation induced by Eq.~(\ref{Eq:BSHamiltonian}) and subsequent detection of leaking photons generates the output state of the mechanical oscillator as
	\begin{equation}
		{\rho _B}({\tau _1}) = \frac{{b{\rho _B}(0){b^\dag }}}{{{\rm Tr}[b{\rho _B}(0){b^\dag }]}} + o ( g \tau_1).
	\end{equation}
	The probability $p_s$ to obtain this state, that is the heralding probability, increases with the effective gain of the optomechanical interaction ($p_s \propto g \tau_1$).
	However, increasing this probability also increases the contribution of multiphoton components in the heralded state, and thus decreases the accuracy of the phonon subtraction.
	Therefore, a weak coupling ($g \tau_1 \ll 1$) is advantageous to realize an approximate phonon subtraction; additionally, the weak coupling reduces heating by photon-recoil.

	In contrast, phonon addition is achieved through a two-mode squeezing interaction,
	\begin{equation}\label{Eq:PSHamiltonian}
		{H_{{\rm{PA}}}} = -g({a^\dag }{b^\dag } + ab)\,,
	\end{equation}
	implementable by driving on the upper mechanical sideband ($\Delta=-\omega_m$)~\cite{aspelmeyer_cavity_2014}.
	The transformation induced by Eq.~(\ref{Eq:PSHamiltonian}) and the subsequent photon detection puts the mechanical mode in a single-phonon-added state described by
	\begin{equation}
		{\rho _B}({\tau _1}) = \frac{{b^\dag{\rho _B}(0){b }}}{{{\rm Tr}[b{\rho _B}(0)b^\dag ]}} + o (g \tau_1).
	\end{equation}

	Although conceptually simple, practical realizations face challenges such as thermal decoherence of the mechanical mode and optical losses, as schematically shown in Fig.~\ref{Fig1}~(b).
	Careful parameter tuning is necessary to mitigate these effects.
	\cref{Eq:BSHamiltonian,Eq:PSHamiltonian} are derived from~\cref{Eq:Lin_Hamiltonian} using rotating-wave approximation that requires the resolved sideband regime where the cavity lifetime is much larger than the mechanical oscillation period ($\kappa^{-1}\gg \omega_m^{-1}$).
	Furthermore, it is beneficial to maintain weak coupling $g \ll \kappa$ and a pulse duration that exceeds the lifetime of the cavity, i.e., $\tau \gg \kappa^{-1}$.
	In this regime, the cavity mode remains approximately unpopulated.
	The photons used for heralding are instead scattered directly to the detector.
	Therefore, the Hamiltonians~(\ref{Eq:PSHamiltonian},\ref{Eq:BSHamiltonian}) accurately describe the interaction between the mechanical oscillator and the light at the detector.
	Incoherent phonon exchange of the mechanical oscillator with its environment including recoil heating (at the rate $\Gamma_m \approx \gamma_m \bar n$, where $\gamma_m$ is the linewidth and $\bar n$ is the mean occupation of the environment) should be negligible.
	Practically it should be slower than the coherent interaction rate $g$ and other rates within the system.

	The initial quantum state of the NP $\rho_B(0)$ that we consider here is either a thermal state $\rho\s{th} (n_0) = \sum_m n_0^m / (1 + n_0)^{m+1} \projector{m}$ with mean occupation $n_0$, or a squeezed thermal state $S(r,\phi) \rho\s{th} (n_0) S^\dagger (r, \phi)$ where $S(r,\phi) = \exp[{ ( \zeta^* a^2 - \zeta (a^\dagger)^2 )/2 }]$ is the squeeze operator parametrized by the squeezing parameter $\zeta = r \exp(i \phi)$.
	In recent literature occupations  as low as $n_0<\num{0.5}$ have been achieved~\cite{delic_cooling_2020,piotrowski_simultaneous_2023}.
	Squeezing of such a thermal state is possible by manipulation with the trapping potential~\cite{janszky_strong_1992,bonvin_state_2023,duchan_experimental_2024} with \SI{4}{\decibel} of squeezing shown in Ref.~\cite{duchan_experimental_2024}.

	To verify the non-Gaussianity of the heralded mechanical state, a red-detuned probe pulse is sent into the cavity.
	The pulse implements a beamsplitter-like interaction akin to~\cref{Eq:BSHamiltonian} and swaps the mechanical state to the field leaking from the cavity.
	By measuring the probe photon statistics, we can reconstruct the mechanical oscillator's phonon statistics and verify that it exhibits quantum non-Gaussianity.
		Non-Gaussianity of the mechanical state can be verified by applying a hierarchy of absolute criteria from Ref.~\cite{lachman_faithful_2019}.
	It accidentally coincides with the modified stellar approach towards photon detection~\cite{chabaud_certification_2021}.
	We define the genuine $n$-phonon quantum non-Gaussianity of the mechanical oscillator as the	phononic states that reject all the mixtures of the states of the form
	\begin{equation}
		\left| {{\Psi _{n - 1}}} \right\rangle  = D(\alpha )S(\xi )\sum\limits_{k = 0}^{n - 1} {{c_k}\left| k \right\rangle }
	\end{equation}
	where $D(\alpha)$ and $S(\xi)$ are the displacement and the squeezing operators.
	The criterion relies on surpassing the thresholds
	\begin{equation}\label{Eq:NG}
		Q_n^G = \mathop {\max }\limits_{\left| {{\Psi _{n - 1}}} \right\rangle } \left\{ {{{\left| {\left\langle n \right.\left| {{\Psi _{n - 1}}} \right\rangle } \right|}^2}} \right\}
	\end{equation}
	i.e. genuine $n$-phonon quantum non-Gaussianity manifests itself when ${Q_n} >Q_n^G$. In Table~\ref{Tab1}, we present the global maximum of $Q_n$ over all pure Gaussian states.
	\begin{table}
		\caption{Maximal probabilities for the genuine $n$-phonon quantum non-Gaussianity.}
		\label{Tab1}
		\begin{tabular}{llllll}
			\hline
			\hline
			$n$     & 1      & 2      & 3      & 4      & 5      \\
			\hline
			$Q_n^G$ & 0.4779 & 0.5574 & 0.5926 & 0.6125 & 0.6249 \\
			\hline
			\hline
		\end{tabular}
	\end{table}
	Our approach provides a pathway for creating more complex nonclassical states of macroscopic mechanical elements using pulsed quantum interactions and measurements.


	\subsection{Experimental Considerations}
While our theoretical framework for phonon addition/subtraction in pulsed optomechanics offers promising insights, translating these concepts to experimental implementation requires careful consideration of practical challenges. In this section, we assess the feasibility of realizing our proposed scheme using state-of-the-art levitated NP systems. We identify key experimental parameters, potential technical obstacles, and strategies to overcome them, demonstrating that our approach, though demanding, lies within reach of current experimental capabilities with appropriate modifications and technical considerations.
Although optical trapping implementation differs significantly between cavity-based and free-space systems, preventing direct performance comparisons, we can evaluate feasibility using parameters from recent experimental demonstrations. Tables \ref{tab:cavity_params} and \ref{tab:freespace_params} list characteristic parameters from state-of-the-art experiments in cavity-based \cite{delic_cooling_2020} and free-space \cite{magrini_squeezed_2022,militaru_ponderomotive_2022} optical levitation systems, respectively.

In the cavity-based implementation, the NP is initially cooled close to its ground state using well-established techniques such as sideband cooling inside a high-finesse cavity \cite{delic_cooling_2020}. During this cooling phase, photons scattered toward the detector are discarded, ensuring that the NP is prepared in a low-occupancy thermal state ($n_0 \ll 1$). At time $t = 0$, the phonon addition/subtraction protocol starts, and the setup is switched to monitor the tweezer photons scattered towards the detector. Phonon subtraction is performed without altering the tweezer detuning—thus maintaining the cooling configuration—whereas phonon addition is realized by applying a pulsed drive on the upper cavity sideband. This process requires a rapid change in the tweezer detuning, which is feasible using electro-optical modulators and is modeled as an instantaneous switch in our theoretical description. Importantly, the coherent-scattering mechanism inherent to the cavity-based approach effectively suppresses elastically scattered tweezer photons from entering the cavity mode, thereby reducing false heralding events. Issues such as the finite spatial extent of the NP, its three-dimensional motion in the tweezer trap, and the finite cavity linewidth can lead to a non-negligible contribution of stray (elastically scattered) photons, which may cause false heralding. An order-of-magnitude estimate, for instance, based on the ratio $g/(\Gamma \times \text{Finesse})$, helps in quantifying the level of filtering required to keep stray light within acceptable limits.

In the free-space configuration, the NP is similarly pre-cooled to near its ground state using techniques such as feedback cooling \cite{kamba_optical_2022,vijayan_scalable_2023}. Unlike the cavity-based system, the free-space setup does not benefit from the inherent rejection of elastically scattered tweezer photons. Instead, it exploits the full quantum nondemolition (QND) interaction to accommodate both elastic and inelastic scattering processes. Here, the pulsed interaction is implemented by selecting a specific temporal window during which scattered photons are treated as heralding events, without the need for rapid detuning changes. The dynamics—including decoherence effects such as recoil heating—are modeled using QLE, with numerical integration employed to accurately capture the system behavior when simplifying approximations (such as the RWA) are not justified. 

We should mention that the method avoids the need for a strong coupling regime.  Although, the mechanical frequencies of the levitated NPs motion are indeed low.
Nevertheless, quantum information protocols are being actively developed for this platform experimentally and theoretically~\cite{neumeier_fast_2024,carlonzambon_motional_2025}.
Moreover, imperfections such as detector inefficiency or optical losses in transmission mainly reduce the photon heralding rate, thereby affecting the phonon addition/subtraction success rate.
However, the heralded mechanical state remains unaffected by optical loss. Mitigation strategies include optimizing pulse durations and interaction strengths, implementing advanced ground-state cooling techniques, and employing high-efficiency photon detectors. These measures collectively ensure that decoherence and recoil heating are sufficiently minimized, thereby enhancing the fidelity of the heralded non-Gaussian mechanical states.

	\begin{table}
		\centering
		\caption{Cavity-based system parameters}
		\label{tab:cavity_params}
		\begin{tabular}{llll}
			\hline\hline
                       & Experimental          & Dimensionless & Simulations \\
			\hline
			$\omega_m$       & $2\pi \times 190$ kHz & 1.96          & 0.05--10    \\
			$\kappa$         & $2\pi \times 96$ kHz  & 1             & 1           \\
			$\gamma \bar{n}$ & $2\pi \times 6$ kHz   & 0.06          & 0--0.06     \\
			$g$              & $2\pi \times 60$ kHz  & 0.62          & 0.02        \\
			\hline\hline
		\end{tabular}
	\end{table}

	\begin{table}
		\centering
		\caption{Free-space system parameters}
		\label{tab:freespace_params}
		\begin{tabular}{llll}
			\hline\hline
                  & Experimental            & Dimensionless & Simulations   \\
			\hline
			 $\omega_m$ & $2\pi \times 73.25$ kHz & 1             & 0.05--10      \\
			 $\gamma$   & $2\pi \times 40$ Hz     & 0.0054        & 0.0054--0.054 \\
			 $\Gamma$   & $2\pi \times 5$ kHz     & 0.068         & 0.0082        \\
			\hline\hline
		\end{tabular}
	\end{table}



	\subsection{Quantum non-Gaussian mechanical states}
	\begin{figure}
		\includegraphics[width=\columnwidth]{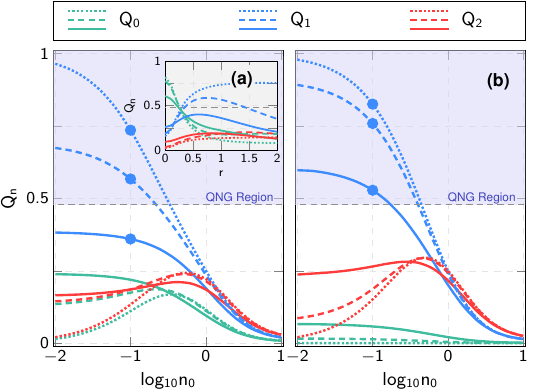}
		\caption{
			Single-phonon-subtracted squeezed thermal mechanical state~(a) and single-phonon-added thermal mechanical state~(b).
			The probability of finding $n$ phonons in the mechanical mode conditioned on detecting a photon at the output versus initial mechanical occupation with initial squeezing with $(r,\phi_0)=(1,\pi/2)$ and (inset) versus squeezing and fixed value of initial mechanical occupation $n_0=0.1$.
			Dotted, dashed, and solid lines correspond to a fixed heating rate $\gamma \bar n= 0$, $\gamma \bar n= 0.01 \kappa$ and  $\gamma \bar n =0.06 \kappa$, respectively.
			The initial state of the mechanics is a squeezed thermal state with mean phonon number $n_0$ and squeezing parameters $(r,\phi_0)$ (\cref{eq:sqz:definitions}) in (a) and a thermal state with mean occupation $n_0$ in (b).
			Other parameters are $g/\kappa=0.02$ and $\kappa \tau=2$.
			The dashed gray line is the first-order absolute quantum non-Gaussianity threshold ${Q_1^G} \approx 0.48$ (\cref{Eq:NG}).
			States marked by the dots are further investigated in Figs.~\ref{Fig4} and \ref{Fig5}.
		}
		\label{Fig2}
	\end{figure}

	Fig.~\ref{Fig2} (a) shows the multiphonon probabilities $Q_n = \mel{n}{\rho_B}{n}$ for the conditional mechanical state $\rho_B$, versus the initial thermal occupation, for the single-phonon subtraction (red-detuned laser pulse), starting from a squeezed thermal state.
	It is known~\cite{kim_nonclassicality_2005} that the subtraction of photons from a thermal state produces a classical output, therefore, for phonon-subtraction, we consider a squeezed thermal initial state.
	In general, subtracting a phonon from a thermal state doubles the occupation of the state~\cite{enzian_singlephonon_2021}, however, for low initial occupations, this effect is insufficient to produce non-Gaussian features, let alone the QNG ones.
	Starting from a squeezed state ${\left| \psi (0) \right\rangle _{m}} \approx \sum_m q_{2m}(r) \left| {2m} \right\rangle $ with $q_{2m}(r)= {\left( {- 1} \right)^m}\sqrt {(2m)!
	} {\tanh ^m}r/[ {{2^{2m}}{{(m!)}^2}\sqrt {\cosh r} }]$,
	a red-detuned pulse can generate a non-Gaussian state since $q_{2}(r)\neq0$.
	Therefore, a QNG state can be achieved with a red-detuned pulse assisted by initial mechanical squeezing, where the single phonon probability $Q_1$ is greater than the maximum probability for Gaussian states (see Table~\ref{Tab1}).
	Due to the squeezed nature of the state, a witness constructed from the Fock state probabilities is intuitively expected to be suboptimal.
	The performance of our witness can be significantly enhanced by extracting the highly nonclassical core of the state via the anti-squeezing operator, $\rho_ B^S = S^\dagger(r)\rho_B S(r)$~\cite{menzies_gaussianoptimized_2009,jezek_experimental_2012}.
	We should mention that squeezing is a Gaussian operation, meaning it transforms mixtures of Gaussian states into other mixtures of Gaussian states.
	Consequently, if $\rho_B^S$ exhibits quantum non-Gaussianity, then the original state $\rho_B$ must also be QNG in nature.
	Therefore, in dealing with the squeezed states, we need to calculate the diagonal density matrix elements in the basis of squeezed Fock states, that is, ${Q_{{n_1}}} \to {Q_{{n_1}}} = \langle {n_1}|{S^\dag }{{\tilde \rho }_B}S|{n_1}\rangle $.
	The effect of squeezing on different phonon probabilities is presented in the inset of Fig~\ref{Fig2}~(a).
	We can see increasing the squeezing parameter up to a certain value allows us to reach single phonon non-Gaussianity.
	After this value, higher Fock states will be populated.
	The squeezing phase does not affect the non-Gaussianity of the generated state and only causes a rotation of the corresponding Wigner function in the phase space.
	Comparing dashed and solid lines in all these Figures we find that the less the heating rate the higher the quantum non-Gaussianity.

	Fig.~\ref{Fig2}~(b) corresponds to the single phonon addition (performed via a pulse with blue detuning).
	A blue-detuned pulse adds a phonon to the state and leaves the system in the conditional state ${\left| \psi(\tau) \right\rangle _m} \approx \left| 1 \right\rangle $ (i.e., $Q_1\approx1$).
	The non-Gaussianity decreases with increasing thermal phonons.
	However, for a blue-detuned pulse with squeezing, the quantum non-Gaussianity is reduced and less resilient to the thermal phonons.
	We note that the multi-phonon probabilities do not imply higher-order non-Gaussianity, as they are all smaller than the maximal probabilities for Gaussian states.

	We next investigate conditionally generated non-Gaussian phonon states and explore how the QND-type interaction necessitates a meticulously precooled mechanical state to approximate phonon addition in a cavityless system.
	As shown in Fig.~\ref{Fig3}, the optomechanical interaction in free space leads to not exactly a phonon addition, unless the mechanical mode is precooled close to the ground state.
	We can see that the scheme is quite robust to the added thermal noise.
	Given the initial cooling to a sufficiently pure state $(n_0\ll 1)$, the heralding approaches a single-phonon addition.
	As the occupation increases, the first-order quantum non-Gaussianity of the state reduces. Fig.~\ref{Fig3}~(b) illustrates the probabilities as a function of pulse duration, revealing a damped oscillatory behavior and highlighting the necessity of employing short pulses to generate a non-Gaussian state.

    As mentioned, the success probabilities of heralding are proportional to the effective strength of the optomechanical interaction $p_s \propto g \tau_1$, and thus, to a certain extent, can be controlled.
    Increasing $p_s$ will also increase the repetition rate of the protocol, at the cost of additional multiphonon contributions of the heralded state.
    Here, we chose these probabilities to be of the order $10^{-3}$, comparable with optical experiments on photon subtraction~\cite{zavatta_subtracting_2008}.

	\newcommand*{\heatingfree}{H_0}
	\begin{figure}
		\includegraphics[width=\columnwidth]{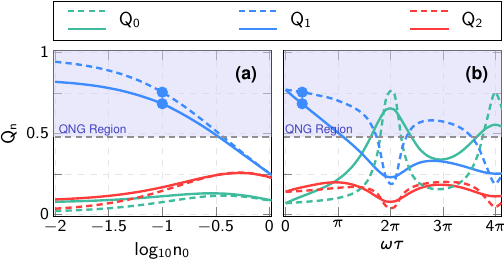}
		\caption{Quantum non-Gaussianity after heralding without a cavity.
			The probability of finding $n$ phonons in the mechanical mode conditioning on detecting photon at the output (a) versus initial mechanical occupation with $\omega \tau=1$ and $\Gamma/\omega_m=0.0082$~\cite{militaru_ponderomotive_2022}.
			(b) Phonon probabilities as a function of the pulse duration assuming $n_0=0.1$.
			Dashed, and solid lines correspond to a fixed heating rate $\gamma \bar n = 10^{-1} \heatingfree$ and $\gamma \bar n = \heatingfree$ respectively, where $\heatingfree = 0.054 \omega_m$.
			The dashed gray line is the first-order absolute quantum non-Gaussianity threshold ${Q_1^G} \approx 0.48$ (\cref{Eq:NG}).
			States marked by the dots are further investigated in Figs.~\ref{Fig4} and \ref{Fig5}. }
		\label{Fig3}
	\end{figure}


	\begin{figure}
		\includegraphics[width=\columnwidth]{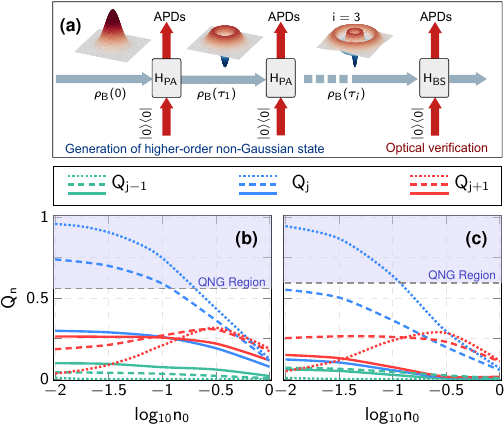}
		\caption{Multiple phonon addition.
			(a) Illustration of the proposed method for generating higher-order QNG mechanical states.
			The probability of finding $n$ phonons in the mechanical mode conditioning on detecting photons at the output versus optomechanical thermal occupation for pulses with blue detuning (b) two pulses $j=2$ and (c) three pulses $j=3$.
			Here $r=0$, $g/\kappa=0.02$ and $\kappa \tau=2$.
			The mechanical mode is initially in equilibrium with the environment $\bar n$.
			Dotted, dashed and solid lines correspond to a fixed heating rate $\gamma \bar n=0 $, $\gamma \bar n= 0.01 \kappa$ and $\gamma \bar n =0.06 \kappa$.
			The dotted black line is the $j$th-order non-Gaussianity threshold.
		}
		\label{Fig6}
	\end{figure}
	As shown schematically in Fig.~\ref{Fig6}~(a), after detecting the photon at the output, we can apply multiple pulses to obtain higher-order quantum non-Gaussianity. Our generated mechanical state after each pulse is not a Gaussian state, however, we can write it in terms of Gaussian states according to
	\begin{equation}
		{\rho _B}({\tau _1}) = \sum\limits_{i = 1}^N {{f_i}\rho _G^i(\tau _1)}\,, \label{Eq:SumDM}
	\end{equation}
	where $\rho_G^i(\tau)$ is a thermal (or squeezed thermal) state with covariance matrix ${\bf{V}}_i$ and $f_i$s are weighting numbers determined numerically. We should note that Eq.~(\ref{Eq:SumDM}) applies to the conditional measurement scheme described here, and may not be generalizable to all possible non-Gaussian states (e.g., states with non-zero first quadrature moments). This representation is a direct consequence of the binary nature of the on-off detector and the fact that one of these outcomes (no click) corresponds to a Gaussian operation while the other (click) gives us our desired non-Gaussian state. This decomposition allows us to write the Wigner function of the system after interaction with the multiple pulses as (a generalized version of Eq.~(\ref{eq:Wigner2}))
	\begin{equation}
		W({\alpha _1},{\alpha _2};{\rho _{AB}}) = \frac{4}{{{\pi ^2}}}\sum\limits_{i = 1}^{N} {\frac{{{f_i}}}{{\sqrt {{{\det \bf{V}}_i}} }}{e^{ - \frac{1}{2}{{\bf{u}}^\dag }{\bf{V}}_i^{ - 1}{\bf{u}}}}},
	\end{equation}
	which means that each state $\rho_G^i(\tau_1)$ undergoes a linear map. We can finally calculate the Wigner function of the mechanical mode after $N$ pulse according to
	\begin{equation}
		{W_B}({\tau _1}) = \sum\limits_{i = 1}^N {{f_i}W_{B,i}(\tau_1)}\,,
	\end{equation}
	In~\cref{Fig6}, we plot multiphonon probabilities after applying second and third pulse versus initial mechanical thermal phonons for two values of heating rate.
	One can easily see that it is possible to achieve higher-order non-Gaussianity if the mechanical mode is cooled close to the ground state.
	Increasing both the initial thermal occupation and the heating rate reduces the effectiveness of our scheme.
	Since the squeezing of the initial mechanical state does not improve the non-Gaussianity, it would be enough to consider the initial thermal state without squeezing.
	An alternative approach to the generation of multiphonon-added states, instead of using multiple pulses each individually implementing a single phonon addition, could be heralding upon multiple photon detection at the optical output.
	Indeed, in such a case, given an optical vacuum at the input, each scattered photon corresponds to a single-phonon addition.
	Unfortunately, the scattering of multiple photons is an event with a proportionally lower probability than scattering a single photon.
	Consequently, to realistically evaluate the outcomes of such events it might be necessary to take into account dark counts of the detectors and other detrimental factors.

		\begin{figure}
		\centering
		\includegraphics[width=\columnwidth]{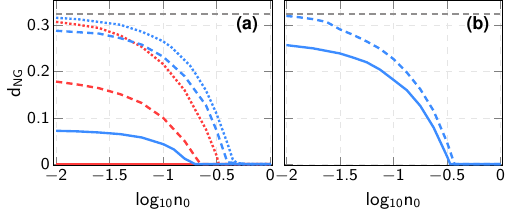}
		\caption{
			Non-Gaussian depth as a function of initial thermal occupation of the mechanical oscillator $n_0$. (a) Phonon-added (blue lines) and subtracted (red lines) and (b) cavityless system. Each curve corresponds to a different mechanical state shown with markers in  Fig.~\ref{Fig2} and Fig.~\ref{Fig3}. The upper bound corresponds to a perfect single phonon state shown by the dashed gray line.
		}
		\label{Fig5}
	\end{figure}

An essential property of QNG states of mechanical motion is their \emph{QNG depth}, which quantifies their resilience against relevant imperfections. Specifically, the QNG depth measures the thermal excitation required to erase quantum non-Gaussianity, as described by Eq.~(\ref{Eq:NG}). For each prepared mechanical state, this metric serves two purposes: (i) assessing the quality of the generated QNG state and (ii) evaluating its robustness against thermalization due to interactions with a thermal bath. For mechanical oscillators, such thermalization is a dominant and well-characterized process thatleads to a gradual loss of QNG features. This process is governed by three key parameters: the damping rate $\gamma$, the thermal occupation number of the environment $\bar{n}$, and the interaction duration $t$. Accordingly, we define the parameter $d_{\text{NG}} = \gamma \bar{n} t$ as a measure of the QNG depth. Physically, $d_{\text{NG}}$ equals the number of thermal quanta absorbed by the NP before its QNG properties vanish. The QNG depth is computed numerically from the Lindblad equation (see methods), which describes the heating process induced by the interaction with a thermal bath. This formulation aligns with the standard framework of quantum open systems, reflecting the expected energy transfer from the bath to the NP. Specifically, $d_{\text{NG}}$ corresponds to the average number of thermal quanta exchanged with the bath during the interaction time $t$. The loss of quantum non-Gaussianity is directly associated with the increase in the thermal phonons of mechanical state, which gradually diminishes its distinct QNG characteristics. By defining $d_{\text{NG}}$ in this manner, we can quantify the amount of heating required for the transition from the QNG regime to a Gaussian or classical-like regime. Experimentally, the critical value of $d_{\text{NG}}$ can be determined by monitoring the state dynamics during a controlled thermalization and identifying the threshold at which the QNG witness is no longer satisfied. This interpretation provides a robust and physically meaningful metric for characterizing the system. Fig.~\ref{Fig5} illustrates the dependence of the QNG depth on the initial mechanical occupation for various mechanical states in Figs.~\ref{Fig2} and \ref{Fig3}. As $n_0$ increases, the system's ability to sustain non-Gaussian characteristics diminishes, illustrating the enhanced thermalization effects at higher $n_0$, which lead to the transition to Gaussian states. Notably, the QNG depth of a perfect single-phonon state $\rho_B=\left| 1 \right\rangle \left\langle 1 \right|$ is $d_{\rm NG}=0.324$, indicated by a dashed gray line. These results highlight the sensitivity of QNG depth to thermal fluctuations and emphasize the necessity of precise control over initial mechanical states to preserve the non-Gaussian features of the heralded states.

	Furthermore, in the supplementary material, we provide a detailed analysis of the measurement and verification scheme for the generated QNG mechanical states.
	This includes the readout process using a second optical pulse with beam-splitter Hamiltonian of Eq.~(\ref{Eq:BSHamiltonian}), the calculation of multiphoton probabilities of the optical pulse state, and the impact of mechanical heating and detection inefficiencies on the transfer of quantum non-Gaussianity from the mechanical mode to the optical one.
	We also present the results of our numerical simulations, highlighting the challenges and optimization strategies for achieving high non-Gaussianity in the state transfer process.

		\begin{figure}
		\centering
		\includegraphics[width=\columnwidth]{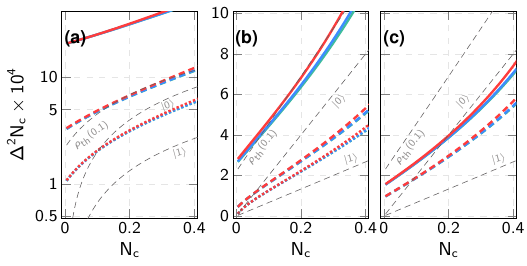}
		\caption{
			Estimation error of phase-randomized force using the heralded mechanical state.
			(a) Photon-subtracted squeezed thermal state, (b) phonon-added thermal state and (c) cavityless optomechanics for the state marked in Figs.~\ref{Fig2} and \ref{Fig3}.
			Different colors correspond to detection capable of resolving different numbers of Fock states (see text): $N=2$ for red, $N=3$ for blue, and $N=4$ for green, assuming $M = 500$ copies of the probe state.
			Solid, dashed, and dotted lines correspond to the same heating rates as in~\cref{Fig2,Fig3}.
			The labeled gray lines show the estimation error when using the vacuum, single phonon, and a thermal state with mean occupation $n_0=0.1$ as the probe state.
		}
		\label{Fig4}
	\end{figure}

	By generating QNG states, we can enhance the system's sensitivity to small displacements.
	To illustrate this, we consider the suitability of the heralded mechanical state for detecting a phase-randomized displacement (PRD)~\cite{oh_optical_2020}.
  The sensing capabilities are evaluated via the quantum Fisher information $F(N_c)$ of the probe state which is computed assuming detection capable of resolving a few lowest Fock states.
	The POVM of this detection includes projectors on these Fock states $\{\Pi_k \equiv \projector{k},\ k = 0,\dots, N\}$, and $1 - \sum_{k=0}^{N} \Pi_k$.
	When $N = 0$, the POVM describes a single on-off detector with unity efficiency.
	The minimal mean-square error in estimating $N_c$ by any unbiased estimator is bounded by the quantum Cramér-Rao inequality, $\Delta^2 N_c \geq 1 / (M F(N_c))$ where $M$ represents the number of probe state copies used in the estimation.
	The complete theoretical framework with necessary details is presented in~\cref{sec:methods}.
	In Fig.~\ref{Fig4}, we investigate the estimation of phase-randomized displacement using the quantum states resulting from approximate phonon subtraction (a), addition (b), and the states heralded in a free-space setup (c).
	Our results demonstrate that the approximate phonon subtraction actually decreases the sensing capability of the state.
	In contrast, approximate addition, either by a cavity or by free-space heralding, provides superior performance compared to the initial thermal Gaussian state even if the heralded states do not surpass the absolute non-Gaussianity threshold given by Eq.~(\ref{Eq:NG}).

  Note that the sensitivity provided by ground-state probes (gray dashed lines labeled $\ket{0}$) correspond to the best sensitivity achievable with classical states, i.e.\ standard quantum limit (SQL).
  In contrast, highly quantum non-Gaussian Fock states are proven (see~\cite{wolf_motional_2019}) to be optimal for sensing displacements with unknown phase.
  These two limits, thus, provide natural benchmarks for the heralded mechanical states.

	One can see from~\cref{Fig4} that lower mechanical noise and consequently higher purity of the heralded mechanical state increases its sensitivity to phase-randomized displacement.
	Moreover, for small values of $N_c$, the estimation error exhibits an approximately linear relationship with $N_c$.
	Importantly, to get almost all the advantages of the heralded state, it is sufficient to resolve up to $N=2$ Fock states.
  Increasing the resolution to $N=3$ Fock states only marginally reduces the estimation error, and further increase of $N$ virtually does not reduce it anymore.

	\section{\label{Sec:Sec3} Discussion}

	We have proposed and investigated a protocol capable of producing approximate mechanical Fock states of the center of mass motion of levitated NPs using pulsed optomechanical interactions and nonlinear photon detection, and specified its applicability for force sensing and storage of QNG features.
	This method can be applied to a multitude of other opto- and electro-mechanical platforms~\cite{aspelmeyer_cavity_2014} and allows comparison of them for the necessary QNG experiments.
	We have also characterized the robustness of the generated non-Gaussian state to various sources of noise and decoherence, such as thermal noise, impure initial states, and inefficient photon detection.
	In a cavity-based setup, the beamsplitter-like coupling of Eq.~(\ref{Eq:BSHamiltonian}) enables efficient verification using established optical measurement techniques~\cite{vanner_optomechanical_2015}, photon counting detectors, and specific non-Gaussianity criteria~\cite{lachman_faithful_2019}.
	In contrast, verification is more challenging with free-space-based systems since they do not have this direct state transfer mechanism.
  However, the methodology~\cite{lachman_quantum_2022} for developing the criteria of quantum non-Gaussianity can be used to derive the requirements with any known state transfer and detection mechanism.

	Here, we addressed the simplest case of adding and subtracting phonons to the thermal mechanical states of levitated NPs trapped in harmonic potentials.
	In principle, NPs can enjoy more complicated motional potentials such as cubic, double-well, or even more exotic ones~\cite{siler_diffusing_2018,rakhubovsky_stroboscopic_2021,riera-campeny_wigner_2023,neumeier_fast_2024,roda-llordes_macroscopic_2024,roda-llordes_numerical_2024}.
	Evolution in nonlinear potentials grants access to motional nonlinearity in the sense of continuous-variable distributions of quadratures.
	The combination of such continuous-variable nonlinearities with discrete operations at a single-phonon level holds promise for a more experimentally versatile approach to quantum state engineering.
	Furthermore, levitated NPs have the capability of freely falling when the trap is switched off, which grants access to unitary Gaussian squeezing.
	This operation is known to enhance higher-order nonlinearities~\cite{neumeier_fast_2024}, and its interplay with the phonon-added and -subtracted states is yet to be studied.
	We have considered sequences of multiple heraldings capable of engineering approximate multiphonon-added and -subtracted states.
	The main factor limiting the accuracy of the heralding is the heating by photon recoil~\cite{jain_direct_2016}.
	Reduction of this heating by adjusting the trapping setup~\cite{gonzalez-ballestero_suppressing_2023} can lead the way towards more elaborate sequences of phonon additions and subtractions.
	Furthermore, ultimately, one can think of sequences of alternating periods of free fall, evolutions in quadratic and nonlinear potentials, and heralding by on-off detection of the light scattered off the NP.

  Our analysis shows that single-phonon control of quantum motion is within reach for levitated systems both inside an optical cavity~\cite{delic_cooling_2020} and in free space~\cite{magrini_squeezed_2022,militaru_ponderomotive_2022}.
	Our work lays the foundation for future studies into the generation, manipulation, and characterization of QNG mechanical states in levitating optomechanics, opening up new directions for fundamental research and technological innovation.

	\section{Methods}
	\label{sec:methods}
	\subsection{Cavity-based pulsed dynamics}
	We use the standard machinery of quantum Langevin equations (QLE) to describe the dynamics of the levitated NP in a tweezer. QLE take into account decoherence of the NP's motion caused by various sources, including recoil heating due to the random scattering of tweezer photons.	 By applying the RWA, we selectively isolate the (anti-)Stokes scattering processes. Under a constant optomechanical coupling (i.e., a flat-top driving power profile), the QLE reduce to ordinary differential equations with fluctuating forces, which can be integrated analytically to yield effective input-output relations. In regimes where the RWA is not applicable, we solve the full QLE numerically (see Fig.~S1 in the Supplementary Materials). This procedure ultimately yields the heralded conditional mechanical state as a function of system parameters (e.g., heralding pulse duration, interaction strength, and heating rates). These parameters are then optimized—primarily the pulse duration and interaction strength—to maximize the quantum non-Gaussianity of the heralded state, with the heating rates chosen to reflect those observed in or accessible to current experiments.
	 The dynamical response of the system to external perturbations and noise is encoded in the QLE, which we cast in a compact matrix form. By	defining the vector of fields $\mathbf{r} = [b^\dagger, a^\dagger, b, a]$, its corresponding vector of noises $\mathbf{r}_{\text{in}} = [b_{\text{in}}^\dagger, a_{\text{in}}^\dagger, b_{\text{in}}, a_{\text{in}}]$, the drift matrix $\mathbf{D}^{\pm}$ (where the plus and minus signs correspond to blue and red detunings, respectively), and the diagonal matrix $\mathbf{N} = \text{diag}[\sqrt{2\gamma}, \sqrt{2\kappa}, \sqrt{2\gamma}, \sqrt{2\kappa}]$, we can express the equations of motion in a compact form:
	\begin{equation}
		\frac{d\mathbf{r}}{dt} = \mathbf{D}^{\pm} \mathbf{r} + \mathbf{N} \mathbf{r}_{\rm in}, \label{Eq:EQM_RWA}
	\end{equation}
	The drift matrices for red and blue detuning are respectively given by
	\begin{equation}
		{{\bf{D}}^ - }\!  \!=\! \! \left[ {\begin{array}{*{20}{c}}
				{ - \gamma }&{  ig}&0&0\\
				{  ig}&{ - \kappa }&0&0\\
				0&0&{ - \gamma }&{-ig}\\
				0&0&{-ig}&{ - \kappa }
		\end{array}} \right]\!\!  ,\, {{\bf{D}}^ + }\! \! =\!\!  \left[ {\begin{array}{*{20}{c}}
				{ - \gamma }&0&0&{  ig}\\
				0&{ - \kappa }&{  ig}&0\\
				0&{-ig}&{ - \gamma }&0\\
				{-ig}&0&0&{ - \kappa }
		\end{array}} \right]\! \! .
	\end{equation}
	Here $\kappa$ and $\gamma$ are the cavity and mechanical damping rates, respectively. The input noise operators satisfy the following non-zero correlators:
	\begin{subequations}
		\label{eq:correlations:definitions}
		\begin{align}
			&	\langle {a_{{\rm{in}}}}(t)a_{{\rm{in}}}^\dag (t')\rangle  = \delta (t - t'){\mkern 1mu} ,\\
			&	\langle {b_{{\rm{in}}}}(t)b_{{\rm{in}}}^\dag (t')\rangle  = \left( {\bar n}+1 \right)\delta (t - t'){\mkern 1mu} ,\\
			&	\langle b_{{\rm{in}}}^\dag (t){b_{{\rm{in}}}}(t')\rangle  =  {\bar n}\delta (t - t'){\mkern 1mu} ,
		\end{align}
	\end{subequations}
	where $\bar n$ is the mean phonon number of the thermal mechanical environment.
	The formal solution of Eq.~(\ref{Eq:EQM_RWA}) is given by
	\begin{equation}\label{key}
		{\bf{r}}(t) = {\bf{M}}^{\pm}(t){\bf{r}}(0) + \int\limits_0^t {ds{\bf{M}}^{\pm}(t - s){\bf{N}}{{\bf{r}}_{{\rm{in}}}}(s)}
	\end{equation}
	with ${\bf{M}}^{\pm}(t) = \exp ({\bf{D}}^{\pm}t)$. The analytical expressions for the elements of matrix ${\bf{M}}^{\pm}(t)$ are given in the Supplemental Material.
	Moreover, the standard input-output theory  $a_{\rm out}(t)= \sqrt {2\kappa} a (t)-a_{\rm in} (t)$ relates the cavity field to the field coming out of the cavity.

	We then define the input and output optical temporal modes~\cite{rakhubovsky_nonclassical_2019}
	\begin{align}
		{\cal A}_{{\rm{in}},{\rm out}}^ \pm (\tau ) = \int\limits_0^\tau  {ds{f_{{\rm{in}},{\rm out}}^ \pm }(s){a_{{\rm{in}},{\rm out}}}(s)} \,,
		\label{Eq:in_out_field}
	\end{align}
	with
	\begin{align}
		{f_{\rm  in}^\pm }(s) &= \sqrt {\frac{{2\kappa }}{{{{{\cal G}^\pm} } - 1}}}\frac{{g{e^{ - \frac{1}{2}(\gamma  + \kappa )(\tau  - s)}}\sinh {G^ \pm }(\tau  - s)}}{{{G^ \pm }}}\,.\\
		{f_{\rm  out}^\pm }(s) &= \sqrt {\frac{{2\kappa }}{{{{{\cal G}^\pm} } - 1}}}\frac{{g{e^{ - \frac{1}{2}(\gamma  + \kappa )s}}\sinh {G^ \pm }s}}{{{G^ \pm }}}\,,
	\end{align}
	and the optomechanical amplification gain $\cal G^\pm$
	\begin{align}
		{{{\cal G}^\pm} } &= 1 + 2\kappa \int\limits_0^\tau  {ds{{\left[ {\frac{1}{{{G^ \pm }}}g{e^{ - \frac{1}{2}(\gamma  + \kappa )s}}\sinh {G^ \pm }s} \right]}^2}} \,.
	\end{align}

	\subsection{Cavityless heralding of levitated phonons}
	The Heisenberg–Langevin equations of motion are then given by
	\begin{align}
		\dot x &= {\omega _m}p\,,\\
		\dot p &=  - {\omega _m}x - \gamma p + \xi (t) + \sqrt {4\Gamma } {X_{{\rm{in}}}}(t)\,,
	\end{align}
	where  the term $\xi (t)$ is the effective thermal force with correlations $\left\langle {\xi (t)\xi (t')} \right\rangle  = 2\gamma \left( {\bar n + 1/2} \right)\delta (t - t')$.
	The NP's motion also influences the quadratures of the interacting modes. We derive the resulting output quadratures using the input-output relation. This establishes the input-output relationship for the optical mode that has a temporally rectangular shape $f_{\rm in,out}=1/\sqrt{\tau}$ as
	\begin{align}
		{\cal X}_{\rm out}(\tau) =&{\cal X}_{\rm in}(\tau)  \\
		{\cal Y}_{\rm out}(\tau) =&{\cal Y}_{\rm in}(\tau)  + \sqrt {4\Gamma/\tau } \int_0^\tau  ds\, x (s)
	\end{align}

	\subsection{Conditional mechanical state based on photon detection}
	We consider the initial mechanical mode to be in a squeezed thermal state with the following correlations
	\begin{subequations}
		\label{eq:sqz:definitions}
		\begin{align}
			&	\langle b(0){b^\dag }(0)\rangle  = \left( {{n_0}\cosh (2r) + {{\cosh }^2}r} \right){\mkern 1mu} ,\\
			&	\langle {b^\dag }(0)b(0)\rangle  = \left( {{n_0}\cosh (2r) + {{\sinh }^2}r} \right){\mkern 1mu} ,\\
			&	\langle b(0)b(0)\rangle  =  - \left( {{n_0} + {1}/{2}} \right){e^{i\phi_0 }}\sinh (2r),
		\end{align}
	\end{subequations}
	where $n_0$ is the mean phonon number of the mechanical thermal state, $r$ is the squeeze parameter and $\phi_0$ is the reference phase for the squeezed mechanical state.
	This mechanical squeezing can be generated through the dynamical instability reached in the far red-detuned and ultrastrong coupling regime in levitated NPs coupled to a microcavity via coherent scattering~\cite{kustura_mechanical_2022} or optically levitated nanosphere by fast switching between two trapping frequencies~\cite{rashid_experimental_2016,bonvin_state_2023}.
	The cavity mode is initially in a vacuum state.
	Mean phonon number of the NP can be reduced $n_0<1$ using active feedback~\cite{gieseler_subkelvin_2012,li_millikelvin_2011,tebbenjohanns_feedback_2023} or coherent scattering~\cite{windey_cavitybased_2019,delic_cavity_2019,delic_cooling_2020,gonzalez-ballestero_theory_2019}.
	Furthermore, strong coupling between the NP and cavity is essential to produce quantum states.
	In these cooling schemes, the NP's damping rate is increased while its average phonon occupation number is reduced to the ground state leaving the mechanical mode in a thermal state with low occupation.
	In addition to the optically induced damping rate, heating from surrounding air molecules and the position fluctuation of the optical tweezer are the limitations of achieving efficient 3D cooling~\cite{gonzalez-ballestero_theory_2019}.
	Specifically, we use parameters similar to recent work demonstrating ground state cooling of a levitated NP through coherent scattering~\cite{delic_cooling_2020}.
	We then apply a laser pulse to this NP which scatters photons from the tweezer laser beam into the cavity mode, creating an optomechanical coupling between the NP and cavity field.
	This pulse generates and monitors non-Gaussian quantum states of the mechanical motion.

	We now define the optomechanical state at time $\tau$ by the vector of fields ${\boldsymbol{u}}^ \pm (\tau ) =({\boldsymbol{a}}^{\dag}, {\boldsymbol{a}} )^T= {[ {{b^\dag }(\tau ),( {\cal A}_{\rm out}^{\pm }(\tau ) )^\dag,b(\tau ),{\cal A}_{\rm out}^ \pm (\tau )} ]^T}$.
	The vector of quadratures ${\boldsymbol{u}}_q^ \pm (\tau ) =({\boldsymbol{p}}, {\boldsymbol{x}} )^T= {[ {{p}(\tau ),{\cal Y}_{\rm out}^{\pm }(\tau ),x(\tau ),{\cal X}_{\rm out}^ \pm (\tau )} ]^T}$ is related to the vector of fields via $\boldsymbol{u}_q^\pm = {\boldsymbol \Omega} \boldsymbol{{u}}^\pm$ where ${\boldsymbol \Omega}$ is a $4\times4$ unitary matrix
	\begin{align}
		\boldsymbol{\Omega} = \frac{1}{{\sqrt 2 }}\left[ {\begin{array}{*{20}{c}}
				{i{{\bf{I}}_2}}&{ - i{{\bf{I}}_2}}\\
				{{{\bf{I}}_2}}&{\;\;{{\bf{I}}_2}}
		\end{array}} \right],
	\end{align}
	with ${\bf I}_2$ being $2 \times 2$ identity matrix.
	The quantum noises in the system follow a Gaussian distribution and the dynamics have been linearized.
	This allows the state of the quantum fluctuations at later times to be modeled as a bipartite Gaussian state with zero means in continuous variables.
	The state is completely described by a $4\times4$ covariance matrix of the operators defined as
	\begin{equation} \label{Eq:CM_fields}
		{{\bf{V}}_{jk}}(\tau) = \frac{1}{2}\langle {\boldsymbol{u}}_{j}^ \pm(\tau ) {\boldsymbol{u}}_{k}^{ \pm \dag }(\tau) + {\boldsymbol{u}}_{k}^{ \pm \dag }(\tau ){\boldsymbol{u}}_{j}^{ \pm }(\tau )\rangle
	\end{equation}
	We present details on calculations of the covariance matrix in the Supplemental Material. We then relate the Wigner function to the covariance matrix according to
	\begin{eqnarray}\label{eq:Wigner2}
		W({\boldsymbol{u}};{\rho _{AB}}) = \frac{4}{{{\pi ^2}\sqrt {\det( {\bf{V}})} }}\exp ( - \frac{1}{2}{{\boldsymbol{u}}^\dag }{{\bf{V}}^{ - 1}}{\boldsymbol{u}}).
	\end{eqnarray}
	As shown in the Supplementary Material, we arrive at the following expression for the unnormalized mechanical Wigner function if we detect $n_2$ photon at the detector
	\begin{align}\label{eq:Wigner4}
		W_B^{n_2}(\alpha)=&	\frac{2 \, \mathcal{ P}_0}{\pi \, n_2!}
		\frac{e^ { - \boldsymbol{v}^{\dag} {\bf L} \boldsymbol{v} }}{\sqrt{\text{det} ({\bf I}_2 + {\bf X}_2 {\bf R}_{mm} )}}
		\nonumber\\
		\times &
		\big(\frac{\partial^2}{\partial \alpha_2 \partial \beta_2^*} \big)^{n_2}
		e^{\frac{1}{2} \boldsymbol{s}_c^T {\bf A} \boldsymbol{s}_c + \boldsymbol{z}^T \boldsymbol{s}_c} \bigg|_{\boldsymbol{s}_c = {\bf 0}},
	\end{align}
	where we have defined $\boldsymbol{v} = (\alpha^*, \alpha)^T $,  $\boldsymbol{s}_c=(\beta _2^*,{\alpha _2})$ and
	\begin{align}
		{\bf L} =&  ({\bf I}_2 + {\bf X}_2 {\bf R}_{mm})^{-1} ({\bf I}_2 - {\bf X}_2 {\bf R}_{mm}),\nonumber\\
		{\bf A} =& {\bf R}_{cc} - {\bf R}_{cm} ({\bf I}_2 + {\bf X}_2 {\bf R}_{mm} )^{-1} {\bf X}_2 {\bf R}_{mc}, \nonumber\\
		\boldsymbol{z} =&  2\, {\bf R}_{cm} ({\bf I}_2 + {\bf X}_2 {\bf R}_{mm})^{-1} \boldsymbol{v}.
	\end{align}
	Here, $\mathcal{ P}_0$ and matrices ${\bf R}_{ij}$ can be expressed in terms of the covariance matrix elements (see Supplementary Material for more details), ${\bf I}_{2}$ is the identity matrix and ${\bf X}_{2}=\boldsymbol{\Omega}^T\boldsymbol{\Omega}$.
	The unnormalized Wigner function is composed of two distinct components.
	The first component is a Gaussian function of the variable $\boldsymbol{v}$.
	The second component involves partial derivatives of a Gaussian function, evaluated at $\boldsymbol{s}_c=0$.
	This results in a polynomial expression in $\boldsymbol{v}$, where the polynomial order depends on the detected photon number pattern.
	When $n_2=0$, indicating no photons registered by the PNRD, the polynomial becomes trivial and equals one.
	In this case, the unnormalized Wigner function simplifies to a Gaussian distribution, implying the mechanical state is Gaussian.
	By comparing Equation (\ref{eq:Wigner4}) for the general Wigner function with Eq.~(\ref{eq:Wigner2}) for a coherent state Wigner function, the covariance matrix ${\bf V}$ can be determined for the $n_2=0$ case:
	\begin{eqnarray}\label{eq:CMnoPhoton-Single}
		{\bf V}_B (n_2 = 0) = \frac{1}{2} ({\bf I}_2 + {\bf X}_2 {\bf R}_{mm}) ({\bf I}_2 - {\bf X}_2 {\bf R}_{mm})^{-1}.
	\end{eqnarray}
	The ability to decompose the Wigner function into Gaussian and polynomial components provides insight into the nature of the mechanical state.
	Specifically, the vanishing of the polynomial contribution for $n_2=0$ reveals the inherent Gaussianity of the state.
	Furthermore, the equation for the covariance matrix $\bf V$ enables full characterization of this Gaussian state.
	Extending this analysis to non-zero photon numbers $n_2\neq0$ would demonstrate how the detection of additional photons introduces non-Gaussian features into the mechanical state.
	The polynomial contribution to the Wigner function is a signature of emerging nonclassical behavior.
	Its structure and order provide quantitative measures of how nonclassical the state becomes.
	Overall, the dual Gaussian and polynomial nature of the Wigner representation furnishes an intuitive picture of how measurement induces transitions from classical to quantum mechanical states.

	A measurement on the optical part with an outcome corresponding to $\Pi_1(\eta)$ will leave the mechanical part in a non-Gaussian state with the density matrix
	\begin{equation}
		{{\tilde \rho }_B} =  \sum\limits_{{n_2} = 0}^\infty  \left[1- {{{(1 - \eta )}^{{n_2}}}}\right] {{\tilde \rho }_B}({n_2}) ,
	\end{equation}
	We can see that conditional measurement results in the elimination of the vacuum component leading to a non-Gaussian state with a Wigner function
	\begin{equation}
		W_B(\alpha)=  \sum\limits_{{n_2} = 0}^\infty  \left[1- {{{(1 - \eta )}^{{n_2}}}}\right] 	W_B^{n_2}(\alpha) ,
	\end{equation}
	The probability of finding $n_1$ phonons in the mechanical mode conditioning on detecting photon at the output is then given by
	\begin{equation}
		{Q_{{n_1}}} = \langle {n_1}|{{\tilde \rho }_B}|{n_1}\rangle  = 2\pi {\cal N}\int {{d^2}\alpha } {W_{{n_1}}}(\alpha ){W_B}(\alpha ),
	\end{equation}
	where  $	 {\cal N}= \int {{d^2}\alpha } {W_B}(\alpha )$ is the normalization constant and ${W_{{n_1}}}(\alpha )$ is the Wigner function of $n_1$-th mechanical Fock state
	\begin{equation}
		{W_{{n_1}}}(\alpha ) = \frac{1}{\pi }{\left( { - 1} \right)^{{n_1}}}\exp \left( { - {\alpha ^2}/2} \right){L_{{n_1}}}\left( {{\alpha ^2}} \right)\,,
	\end{equation}
	with $L_{n_1}(\alpha)$ denoting the $n_1$-th Laguerre polynomial.

	A phase randomized displacement operation ~\cite{oh_optical_2020} transforms the initial mechanical
	state $\rho_B(\tau)$ into an output state $\rho_{B,f}$ according to
	\begin{align}
		{\rho _{B,f}} &= \int_0^{2\pi } {\frac{{d\phi }}{{2\pi }}} D(\sqrt {{N_c}} {e^{i\phi }}){\rho _B}(\tau ){D^\dag }(\sqrt {{N_c}} {e^{i\phi }})\nonumber\\ &= \sum\limits_{n = 0} {{p_f}(n|{N_c})\left| n \right\rangle \left\langle n \right|},
	\end{align}
	with
	\begin{align}
		{p_f}(n|{N_c}) &= \sum\limits_{m = 0} {{q_m}{{\left| {\left\langle n \right|D(\sqrt {{N_c}} )\left| m \right\rangle } \right|}^2}} \nonumber\\
		&= \sum\limits_{m = 0} {{q_m}\frac{{m!}}{{n!}}{e^{{-N_c}}}N_c^{n - m}L_m^{n - m}{{({N_c})}^2}}
	\end{align}
	To compute the Fisher information of this probe state, we assume detection that can resolve a few lowest Fock states (up to $k_{\max}$).
	Such detection is characterized by the POVM including projectors on the few lowest Fock states: $\Pi_k = \projector{k}$ for $k = 0,\dots,k_{\max}$, and a projector on the rest of the Fock space $\Pi_{k+} = 1 - {\sum_k^{k_{\max}}} \Pi_k$.
	The Fisher information of the probe state corresponding to this POVM is
	\begin{multline}
		F({N_c}) = \sum\limits_{n = 0}^{{k_{\max }}} {\frac{1}{{{p_f}(n|{N_c})}}{{\left[ {\frac{{\partial {p_f}(n|{N_c})}}{{\partial {N_c}}}} \right]}^2}}
		\\
		+ \frac{1}{{{p_{k + }}}}{\left[ {\frac{{\partial {p_{k + }(N_c)}}}{{\partial {N_c}}}} \right]^2}
	\end{multline}
	with
	\begin{equation}
		{p_{k + }}({N_c}) = \sum\limits_{n = {k_{\max }} + 1}^\infty  {{p_f}(n|{N_c})}
	\end{equation}

	To evaluate the QNG and the non-classicality depth, we examine the evolution of the mechanical density matrix given by the master equation:
	\begin{equation}
		\frac{d}{dt} \rho = \frac{\gamma}{2} \left\{ (\bar{n} + 1){\cal L}[b] + \bar{n}{\cal L}[b^\dagger] \right\}
	\end{equation}
	where ${\cal L}[o] = 2o\rho o^\dagger - o^\dagger o\rho - \rho o^\dagger o$ represents the Lindblad superoperator. Subsequently, we transform the master equation into a Fokker-Planck equation for the Wigner function:
	\begin{multline}
		\frac{d}{dt}W(x,y)
		\\
		= \frac{\gamma}{2} \bigg\{ \partial_x x + \partial_y y + \left(\bar{n} + \frac{1}{2}\right)\partial_{xx} + \partial_{yy} \bigg\}W(x,y)
	\end{multline}

\section*{Data availability}
Data sharing not applicable to this article as no datasets were generated or analysed during the current study.

	\begin{acknowledgments}
		We acknowledge the project 23-06308S of the Czech Science Foundation and project CZ.02.01.01/00/22\_008/0004649 of the MEYS Czech Republic supported by the EU funding.
		R.F. also acknowledges funding from the MEYS of the Czech Republic (Grant Agreement 8C22001), Project SPARQL has received funding from the European Union's Horizon 2020 Research and Innovation Programme under Grant Agreement no. 731473 and 101017733 (QuantERA).
	\end{acknowledgments}

\section*{Author contributions}
R.F. conceived the theoretical idea, initiated and coordinated the project, and provided supervision throughout. A.R. collaborated with R.F. in developing the core concepts. F.B. performed both analytical and numerical calculations, with all authors participating in the interpretation of results. The manuscript was primarily written by F.B. and A.R., with significant input from R.F. All authors were actively involved in the creation, revision, and finalization of the manuscript.

\section*{Competing interests}
The authors declare no competing interests.

\title{Supplementary material for ``\thetitle''}

\renewcommand{\thefigure}{S\arabic{figure}}
\renewcommand{\theHfigure}{S\arabic{figure}}
\renewcommand{\theequation}{S\arabic{equation}}
\setcounter{figure}{0}
\setcounter{equation}{0}
\setcounter{section}{0}
\onecolumngrid
\pagebreak
\begin{center}
	\textbf{{\large Supplementary material for ``\thetitle''}}\\
{F. Bemani}, {A. A. Rakhubovsky} and
{R. Filip}\\
{Department of Optics, Palack\'{y} University, 17. listopadu 1192/12, 77146 Olomouc, Czech Republic}
\end{center}

\section{Levitated optomechanics Hamiltonian} 
\label{sec:coherent_scattering_based_coupling}

The cavity-based configuration we consider in this work is assumed to realize the coherent-scattering-based coupling of the center-of-mass motion of the NP and the cavity optical mode.
For simplicity, here we consider the motion of the NP in only one direction.
The interaction Hamiltonian corresponding to this coupling reads~\cite{gonzalez-ballestero_theory_2019}
\begin{equation}
	H / \hbar = \Delta a^\dag a + \frac{ p_m^2 }{2 m } + \frac 12 m \omega_m^2 x_m^2 - G ( a + a^\dag) x_m,
\end{equation}
where $x_m, p_m$ are the position and momentum of the NP ($\comm{x_m}{p_m} = i \hbar$) with mass $m$ and frequency $\omega_m$, $a$ is the annihilation operator of the cavity mode, and $G$ is the optomechanical coupling rate.
The frequency $\omega_m$ is defined by the properties of the trapping light beam:
\begin{equation}
	\omega_m = \sqrt{ \frac{ \epsilon_c }{ \rho W_t^2 } } \sqrt{ \frac{ 4 P_t }{ \pi c \mathcal A }},
\end{equation}
where $c$ is the speed of light in vacuum, $\epsilon_c = 3 ( \epsilon - 1)/(\epsilon + 2)$, $\epsilon$ is the relative permittivity of the NP's material, $\rho$ is the mass density of the NP.
Furthermore, $W_t$ is the tweezer waist, $\mathcal A$ is the tweezer cross-section area in the waist, $P_t$ is the tweezer power.

The coupling rate $G$ is given by
\begin{equation}
	G = \epsilon_c V \sqrt{ \frac{ 4 P_t }{ \pi  c \mathcal{A}}} \sqrt{ \frac{ \omega_c }{ 2 \hbar V_c }}.
\end{equation}
Here 
$V$ is the NP's volume, $\omega_c$ is the cavity frequency, $V_c = \pi W_c^2 L_c / 4$ is the cavity mode volume determined by the cavity waist $W_c$ and length $L_c$.

Without a cavity, the optomechanical coupling in the free space is given by the coupling rate~\cite{magrini_squeezed_2022}
\begin{equation}
	\Gamma = 
	2 \pi  \frac{ \alpha k }{ \hbar } \sqrt{ \frac{ 7 \pi \hbar \omega_m^3 }{ 15 \epsilon_0 ( 2 \pi c)^3 } }\frac{ \hbar }{ 2 m \omega_m } \sqrt{ \frac{ 4 P_t }{ \pi \epsilon_0 c \mathcal{A}}}.
\end{equation}
Here $\alpha = \epsilon_0 \epsilon_c V$ is the NP's polarizability, $\epsilon_0$ is the vacuum permittivity, $k = \omega_c/c$ is the light wave vector.


\section{Time-dependent drift matrix}
\begin{figure}
	\centering
	\includegraphics[scale=1]{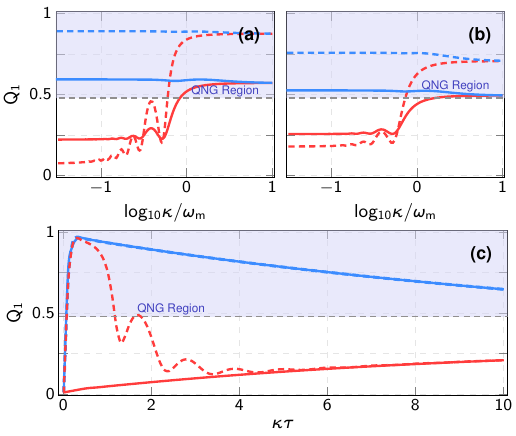}
	\caption{Single phonon probability of the heralded mechanical state as a function of sideband parameter for fixed pulse duration $\kappa \tau=2$.
		Red and blue curves correspond to the red and blue detuning, respectively.
		Dashed lines correspond to a fixed heating rate $\gamma \bar n= 0.01 \kappa$ while the solid lines correspond to $\gamma \bar n =0.06 \kappa$, (a) $n_0=0.01$ and (b) $n_0=0.1$.
		(c) Single phonon probability as a function of pulse duration. $\omega_m=3 \kappa$, $\gamma \bar n= 0.01 \kappa$.
		Dashed lines are numerical solutions without RWA and solid lines are with considering RWA.
		The dotted black line is the absolute threshold of single-phonon quantum non-Gaussianity.
	}
	\label{FigS1}
\end{figure}

In the main text, we have considered the dynamics of the system under the rotating wave approximation (RWA), which simplifies the equations of motion by neglecting the rapidly oscillating terms.
The validity of the RWA rests upon the assumptions of good sideband resolution ($\omega_m \gg \kappa$) and weak optomechanical coupling ($g \ll \kappa,\omega_m$).
Given that a rather strong optomechanical coupling is available with levitated NPs (e.g. $g/\kappa \approx 0.6$ was reached in~\cite{delic_cooling_2020}), here we analyze the optomechanical dynamics beyond RWA.
We introduce a time-dependent optomechanical coupling rate $g(t)$.
This coupling rate modulates the interaction strength between the optical and mechanical modes.
The Hamiltonian of the system in the rotating frame, without invoking the RWA, contains terms oscillating as $\exp[{ \pm 2i{\omega_m}t}]$.
The inclusion of these terms or considering a time-dependent coupling rate leads to a time-dependent drift matrix $\mathbf{D}(t)$.
To include the effect of counter-rotating terms we use the linearized optomechanical Hamiltonian of Eq.~(1).
The equations of motion then arise as
\begin{equation}
	\frac{d\mathbf{r}}{dt} = \mathbf{D}^{\pm} \mathbf{r} + \mathbf{N} \mathbf{r}_{\rm in},
\end{equation}
Since we are interested in the solution for the pulses of the leaking light we take a derivative of Eq.~(17) over time.
That is, we have introduced $\mathbf{r}(t)= {[ {{b^\dag },a^{\dagger},( {\cal A}_{\rm out}^{\pm } )^\dagger (\tau ),b,a,{\cal A}_{\rm out}^ \pm } ]^T}$ as the vector of system operators, and $\mathbf{N} {\mathbf{r}}_{\rm in}$ its corresponding vector of noises.
The drift matrices for blue and red detuning are
\begin{equation}
	{{\bf{D}}^ + }   =
	\begingroup 
	\setlength\arraycolsep{1pt}
	\begin{bmatrix}
		{ - \gamma }&{ - ig{e^{ - 2i{\omega _m}t}}}&0&0&{ - ig}&0\\
		{ - ig{e^{ - 2i{\omega _m}t}}}&{ - \kappa }&0&{ - ig}&0&0\\
		0&{\sqrt {2\kappa } {f_{{\rm{out}}}}(t)}&0&0&0&0\\
		0&{ig}&0&{ - \gamma }&{ig{e^{2i{\omega _m}t}}}&0\\
		{ig}&0&0&{ig{e^{2i{\omega _m}t}}}&{ - \kappa }&0\\
		0&0&0&0&{\sqrt {2\kappa } {f_{{\rm{out}}}}(t)}&0
	\end{bmatrix}
	\endgroup\,,
\end{equation}
\begin{equation}
	{{\bf{D}}^ - }   =
	\begingroup 
	\setlength\arraycolsep{1pt}
	\begin{bmatrix}
		{ - \gamma }&{ - ig}&0&0&{ - ig{e^{ - 2i{\omega _m}t}}}&0\\
		{ - ig}&{ - \kappa }&0&{ - ig{e^{ - 2i{\omega _m}t}}}&0&0\\
		0&{\sqrt {2\kappa } {f_{{\rm{out}}}}(t)}&0&0&0&0\\
		0&{ig{e^{2i{\omega _m}t}}}&0&{ - \gamma }&{ig}&0\\
		{ig{e^{2i{\omega _m}t}}}&0&0&{ig}&{ - \kappa }&0\\
		0&0&0&0&{\sqrt {2\kappa } {f_{{\rm{out}}}}(t)}&0
	\end{bmatrix}
	\endgroup\,,
\end{equation}
with the matrix
\begin{equation}
	\mathbf{N}=\left[ {\begin{array}{*{20}{c}}
			{\sqrt {2\gamma } }&0&0&0&0&0\\
			0&{\sqrt {2\kappa } }&0&0&0&0\\
			0&{ - {f_{{\rm{out}}}}(t)}&0&0&0&0\\
			0&0&0&{\sqrt {2\gamma } }&0&0\\
			0&0&0&0&{\sqrt {2\kappa } }&0\\
			0&0&0&0&{ - {f_{{\rm{out}}}}(t)}&0
	\end{array}} \right]\,.
\end{equation}
The system is then completely described by a $6\times6$ covariance matrix defined as
\begin{equation}
	{{\bf{W}}_{jk}}(\tau) = \frac{1}{2}\langle {\boldsymbol{r}}_{j}(\tau ) {\boldsymbol{r}}_{k}^{ \dag }(\tau) + {\boldsymbol{r}}_{k}^{ \dag }(\tau ){\boldsymbol{r}}_{j}(\tau )\rangle 	\,.
\end{equation}
The evolution of the covariance matrix $\mathbf{W}(t)$ of this three-mode system is governed by the Lyapunov equation~\cite{mari_gently_2009}:
\begin{equation}
	\frac{d}{dt}{\mathbf{W}}(t) = \mathbf{D}(t)\mathbf{W}(t) + \mathbf{W}(t)\mathbf{D}^\dagger(t) + \mathbf{F},
\end{equation}
where $\mathbf{F}$ is the diffusion matrix,
\begin{equation}
	{\bf{F}} = \left[ {\begin{array}{*{20}{c}}
			{{{\bf{0}}_{3 \times 3}}}&{{\bf{\tilde F}}}\\
			{{\bf{\tilde F}}}&{{{\bf{0}}_{3 \times 3}}}
	\end{array}} \right]\,,
	\text{ where }
	{\bf{\tilde F}} = \left[ {\begin{array}{*{20}{c}}
			{\gamma (2\bar n + 1)}&0&0\\
			0&\kappa &{ - \frac{1}{2}\sqrt {2\kappa } {f_{{\rm{out}}}}(t)}\\
			0&{ - \frac{1}{2}\sqrt {2\kappa } {f_{{\rm{out}}}}(t)}&{\frac{1}{2}{f_{{\rm{out}}}}{{(t)}^2}}
	\end{array}} \right]\,,
\end{equation}
accounting for the noise correlations.
To trace out the intracavity mode, we remove the rows and columns associated with mode $a$ from the covariance matrix $\mathbf{W}$.
This leaves us with the $(4 \times 4)$ block $\mathbf{V}$ that corresponds to the mechanical mode and the output optical mode given by Eq.~(27).
Then, we can calculate the desired parameters numerically in the same manner as before.

Fig.~(\ref{FigS1}) illustrates the probability of generating a single phonon state in the mechanical system as a function of the sideband parameter $\kappa/\omega_m$.
The solid and dashed curves correspond to different levels of mechanical heating.
The red (blue) curves are for a red (blue) detuned laser.
For blue detuning, the probability is almost constant.
For the red detuning, the probability of obtaining a single phonon increases as the sideband parameter increases, until it reaches a maximum value.
In the red detuning scenario with a lower heating rate of $\gamma \bar n= 0.01 \kappa$ (dashed curves), the maximum single phonon probability approaches a larger value compared to the case with a higher heating rate of $\gamma \bar n= 0.06 \kappa$ (solid curves).
This highlights the detrimental effects of thermal decoherence.
Beyond the RWA, the interaction between a mechanical mode and an optical mode, which is initially prepared in its vacuum state, can be described by Hamiltonian ${H_{{\rm{BRWA}}}}=-g(a+a^\dagger)(b+b^\dagger)$.
By applying a unitary transformation $U(t) = \exp \left( {igt H_{\rm BRWA}} \right)$, we can obtain the time evolution of the system.
This process can be understood as the action of the unitary transformation on the two input modes, resulting in the creation of the desired state.
That is
\begin{align}
	U(t)&{\rho ^m}(0) \otimes \left| 0 \right\rangle \left\langle 0 \right|{U^\dag }(t)\nonumber\\\
	\approx& {\rho ^m}(0) \otimes \left| 0 \right\rangle \left\langle 0 \right| + {g^2}{t^2} \left[b{\rho ^m}(0){b^\dag }  + {b^\dag }{\rho ^m}(0)b \right] \otimes \left| 1 \right\rangle \left\langle 1 \right| \nonumber\\\
	&+ igt\left( {b{\rho ^m}(0) \otimes \left| 1 \right\rangle \left\langle 0 \right| - {\rho ^m}(0){b^\dag } \otimes \left| 0 \right\rangle \left\langle 1 \right|} \right) \nonumber\\\
	&+ igt\left( {{b^\dag }{\rho ^m}(0) \otimes \left| 1 \right\rangle \left\langle 0 \right| - {\rho ^m}(0)b \otimes \left| 0 \right\rangle \left\langle 1 \right|} \right).
\end{align}
If we detect a photon in the output optical mode, the mechanical state is projected into
\begin{equation}
	{\rho_B}\propto{b{\rho ^m}(0)b^\dag} +{b^\dagger{\rho ^m}(0)b}.
\end{equation}
For very low thermal phonons (as in Fig.~\ref{FigS1}), the initial mechanical state can be approximated as a vacuum state ${\rho ^m}(0) \approx \left| 0 \right\rangle \left\langle 0 \right| $. Therefore, we can neglect the first term, and the conditionally generated mechanical state can be approximated as a phonon-added state.

In Fig.~\ref{FigS1}~(c), we plot the single phonon probability as a function of pulse duration for a fixed value of the sideband ratio. We observe that the results of considering the RWA exactly match those beyond the RWA for blue detuning, as well as for long red-detuned pulses. However, for short red-detuned pulses, they do not match.

Using the same procedure, we have checked that a time-dependent optomechanical coupling rate as $g(t)=\bar{g}[1-\exp (-\kappa t)]$ will not affect the results of the main text.

\section{Details on measurement and verification scheme}
\begin{figure}
	\includegraphics[scale=1]{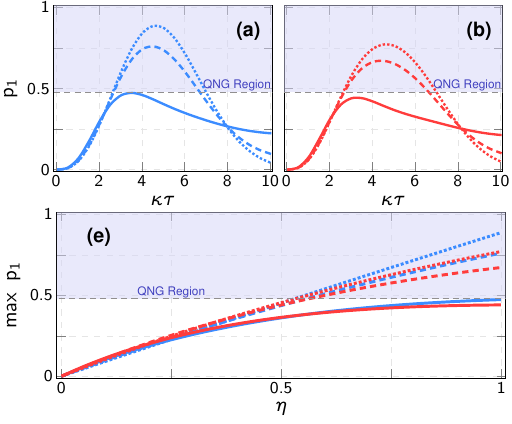}
	\caption{
		Probability of detecting a single photon at the output with unit detection efficiency after applying the readout pulse.
		The mechanical state is a phonon-added state generated from the initial mechanical thermal state with (a) $n_0=0.01$ and (b) $n_0=0.1$.
		The dotted, dashed, and solid lines correspond to a fixed heating rate $\gamma \bar n=0 $, $\gamma \bar n= 0.01 \kappa $, and $\gamma \bar n =0.06 \kappa$, respectively.
		(c) Maximum single photon detection probability versus detection efficiency for the states generated in (a) and (b).
		Here we fix the optomechanical interaction coupling rate $g/\kappa=0.6$.
	}
	\label{FigS2}
\end{figure}
The generated mechanical state is relatively robust against decoherence and noise on timescales comparable to the mechanical decoherence time.
Here, we study the readout of the generated mechanical state by applying another optical pulse as shown in Fig.~1~(c).
By adjusting the parameters of the second laser pulse, the resulting mechanical state can be measured through photon detection at the output.
This process is based on the beam-splitter interaction Hamiltonian of Eq.~(4).
By adjusting the duration $\tau_2$ and optomechanical interaction rate $g$ of the second optical pulse and assuming the mode ${\cal A}_{{\rm{in}}}^-$ is in a vacuum state, then detecting $n$ photons in ${\cal A}_{\rm out}^-$ implies that $n$ phonons have been subtracted from the initial mechanical mode $b(\tau_1)$.
This allows the realization of a phonon subtraction operation on the mechanical state.
It also indicates that, within the chosen range of parameters, the second pulse's output light provides a direct measurement of the mechanical state's behavior.
By determining the probabilities of photon detection, all elements of the mechanical covariance matrix and, consequently, the mechanical density matrix can be identified.
We can obtain the measurement probability $p(n)$ starting from a Gaussian state by performing a trace of the unnormalized density operator $\tilde \rho_B(n)$, which corresponds to integrating the unnormalized Wigner function Eq.~(29).
The integration over $\alpha$ leading to the following photon detection probability
\begin{equation}
	p(n) = \frac{{{{\cal P}_0}}}{{n!{\mkern 1mu} \sqrt {{\rm{det}}({{\bf{I}}_2} - {{\bf{X}}_2}{{\bf{R}}_{mm}})} }}\frac{{{\partial ^{2n}}}}{{\partial \alpha _2^n\partial \beta _2^{*n}}}{e^{\frac{1}{2}{\bf{s}}_c^T{{\bf{A}}_p}{{\bf{s}}_c}}}{|_{{{\bf{s}}_c} = 0}},
\end{equation}
with
\begin{equation}
	{\bf A}_p = {\bf R}_{cc} + {\bf R}_{cm} ({\bf I}_2 - {\bf X}_2 {\bf R}_{mm} )^{-1} {\bf X}_2 {\bf R}_{mc}.
\end{equation}
Since the generated mechanical state is not a Gaussian state, write it in terms of Gaussian states according to Eq.~(10).
This allows us to calculate the Wigner function of the optical mode at the output. We can calculate the probability of detecting a photon according to
\begin{align}
	p(n) =& \frac{{{f_1}{\cal P}_0^1}}{{n!\sqrt {{\rm{det}}({{\bf{I}}_2} - {{\bf{X}}_2}{\bf{R}}_{mm}^1)} }}\frac{{{\partial ^{2n}}}}{{\partial \alpha _2^n\partial \beta _2^{*n}}}{e^{\frac{1}{2}{\bf{s}}_c^T{\bf{A}}_p^1{{\bf{s}}_c}}}{|_{{{\bf{s}}_c} = 0}}\nonumber\\
	&	+ \frac{{{f_2}{\cal P}_0^2}}{{n!\sqrt {{\rm{det}}({{\bf{I}}_2} - {{\bf{X}}_2}{\bf{R}}_{mm}^2)} }}\frac{{{\partial ^{2n}}}}{{\partial \alpha _2^n\partial \beta _2^{*n}}}{e^{\frac{1}{2}{\bf{s}}_c^T{\bf{A}}_p^2{{\bf{s}}_c}}}{|_{{{\bf{s}}_c} = 0}},
\end{align}
where ${\bf R}^i_{mm}$ and ${\bf A}_p^i$ are calculated in terms of covariance matrices ${\bf V}^i$. The effect of detector efficiency can be modeled by inserting a beam splitter of transmissivity $\eta$.
The output field ${\cal A}_{\rm out}$, which contains information from the mechanics, is superimposed on the field from an ancillary mode ${\cal A}_{\rm anc}$ at a lossless beam-splitter with transmissivity $\eta$. Then denoting the two out-modes ${{\cal A}'}_{\rm out}$ and ${{{{\cal A}'}_{\rm anc}}}$, respectively, we have
\begin{align}
	{{\cal A}' }_{\rm out}& = \sqrt \eta  {{\cal A}_{\rm out}} + i\sqrt {1 - \eta } {{\cal A}_{\rm anc}}\,,\\
	{{\cal A}' }_{\rm anc} &= i\sqrt {1 - \eta } {{\cal A}_{\rm out}} + \sqrt \eta  {{\cal A}_{\rm anc}}\,.
\end{align}
This transformation comprises ${{\cal A}'_{\rm out}} = {U^\dag }{{\cal A}_{\rm out}}U$ and ${{\cal A}'_{\rm anc}} = {U^\dag }{\cal A}_{\rm anc}U$ with $U = {e^{i\pi {J_3}/2}}{e^{ - 2i\theta {J_2}/2}}{e^{ - i\pi {J_3}/2}}$ and the operators
\begin{align}
	{J_2} = \frac{1}{{2i}}\left[ {{\cal A}_{\rm anc}^\dag {{\cal A}_{\rm out}} - {\cal A}_{\rm out}^\dag {{\cal A}_{\rm anc}}} \right]\\
	{J_3} = \frac{1}{2}\left[ {{\cal A}_{\rm anc}^\dag {{\cal A}_{\rm anc}}- {\cal A}_{\rm out}^\dag {{\cal A}_{\rm out}}} \right]\,.
\end{align}
We consider the optical state before the beam-splitter $\rho_B$ and the vacuum for the ancillary optical mode. Therefore, the system in the state $\rho=\rho_{\rm out} \otimes {\left| {{0}} \right\rangle \left\langle {{0}} \right|}$ will transform according to $\rho'=U \rho U^\dagger $. The output state after the interaction is given by
\begin{align}
	\rho ' =& \sum\limits_{m,n = 0} {\left[ {\frac{{{e^{ - i(m - n)\pi /2}}}}{{\sqrt {m!n!} }}{{( - 1)}^{m + n}}{{\left( {1 - \eta } \right)}^{\frac{{m + n}}{2}}}{\eta ^{-\frac{{m + n}}{2}}}} \right.} \nonumber\\
	&	\left. { \times {\cal A}_{\rm out}^m{\eta ^{\frac{1}{2}{\cal A}_{\rm out}^\dag {{\cal A}_{\rm out}}}}{\rho _{\rm out}}{\eta ^{\frac{1}{2}{\cal A}_{\rm out}^\dag {{\cal A}_{\rm out}}}}{\cal A}_{\rm out}^{\dag n} \otimes \left| m \right\rangle \langle n|} \right]\,.
\end{align}
We then trace out the ancillary mode $\rho'_{\rm out}= {\rm Tr} _{\rm anc} [\rho]$ and project into the Fock states to obtain different elements of the output density matrix according to
\begin{equation}
	{{p'}_{i,j}} = \sum\limits_{l = 0} {\frac{1}{{l!}}\sqrt {\frac{{\left( {i + l} \right)!\left( {j + l} \right)!}}{{i!j!}}} {{\left( {1 - \eta } \right)}^l}{\eta ^{\frac{{i + j}}{2}}}{p_{i + l,j + l}}} \,.
\end{equation}
Specifically, the diagonal elements which give the photon detection probabilities are given by
\begin{equation}
	{{p'}_{i,i}} = \sum\limits_{l = 0} {\left( \begin{array}{c}
			i + l\\
			l
		\end{array} \right){{\left( {1 - \eta } \right)}^l}{\eta ^i}{p_{i + l,i + l}}} \,.
\end{equation}
The transfer of non-Gaussianity from the mechanical mode to the optical mode is quantified by the non-Gaussianity parameter $p_1$.
The detection schemes for the generated non-Gaussian mechanical states follow similar principles, irrespective of whether the states are phonon-added or phonon-subtracted.
For illustrative purposes, in Fig.~\ref{FigS2} we focus on the phonon-added case, with the understanding that analogous considerations apply to the phonon-subtracted scenario.
Panels (a) and (b) correspond to the probability of detecting a single photon at the output with unit detection efficiency after applying the readout pulse for various different values of mechanical heating rate.
These plots suggest the effects of mechanical heating on the transfer of non-Gaussianity as a key factor.
Comparing the (a) blue and (b) red curves, one finds that the non-Gaussianity is reduced by increasing the initial thermal phonons of the mechanical mode.
This highlights the challenge of maintaining high non-Gaussianity in the presence of thermal phonons (compare red and blue) and heating rate.
Besides, inevitable detection inefficiencies and losses can obscure these non-Gaussian traits, effectively convolving the true photon statistics with a Gaussian noise profile.
The non-Gaussianity parameter is highest at the state swap time, $\tau_s\approx\pi /2| G^-|$, where the mechanical and optical modes exchange their quantum states.
To investigate the effect of inefficiency, in Fig.~\ref{FigS2}~(c), we show the maximum single photon detection probability versus detection efficiency for the states generated in Fig.~2~(b).
Moreover, if the detection efficiency is reduced to below $\eta<0.5$ even in the perfect case shown by dotted lines, the non-Gaussianity of the mechanical mode cannot be transferred to the optical mode.
We should notice that these results are obtained in a different regime of parameters than the generation part, with a longer pulse duration $\kappa \tau \approx 4$ and a stronger optomechanical coupling $g/\kappa=0.6$.
This suggests that the optimal parameters for state generation and state transfer may be different, and careful optimization is required to achieve high non-Gaussianity in both processes.
We note that the insights gleaned from the phonon-added state are extendable to its subtracted counterpart, affirming the symmetry in their non-Gaussian character and detection methodology.

\section{\label{Sec:App1} Elements of the matrix $\bf {M}^{\pm}$}
The nonzero elements of the matrix $\bf {M}^{\pm}$ are given by
\begin{equation}
	{\bf{M}}_{11}^ \pm  = {\bf{M}}_{33}^ \pm  = {e^{ - \frac{1}{2}\left( {\gamma  + \kappa } \right)\tau }}\left[ {\cosh {G^ \pm }\tau  + \frac{{\kappa  - \gamma }}{{{G^ \pm }}}\sinh {G^ \pm }\tau } \right]
\end{equation}
\begin{equation}
	{\bf{M}}_{22}^ \pm  = {\bf{M}}_{44}^ \pm  = {e^{ - \frac{1}{2}\left( {\gamma  + \kappa } \right)\tau }}\left[ {\cosh {G^ \pm }\tau  - \frac{{\kappa  - \gamma }}{{{G^ \pm }}}\sinh {G^ \pm }\tau } \right]
\end{equation}
\begin{equation}
	-{\bf{M}}_{21}^ -  = -{\bf{M}}_{12}^ -  =   {\bf{M}}_{34}^ -  =   {\bf{M}}_{43}^ -  =   \frac{{ig} {e^{ - \frac{1}{2}\left( {\gamma  + \kappa } \right)\tau }}}{{{G^ - }}}\sinh {G^ - }\tau
\end{equation}
\begin{equation}
	-{\bf{M}}_{14}^ +  =   {\bf{M}}_{41}^ +  =- {\bf{M}}_{23}^ +  =   {\bf{M}}_{32}^ +  =   \frac{{ig{e^{ - \frac{1}{2}\left( {\gamma  + \kappa } \right)\tau }}}}{{{G^ + }}}\sinh {G^ + }\tau
\end{equation}
with $	{G^ \pm } = \sqrt {\pm g^2 + {(\kappa -\gamma)^2}/4} $.

\section{\label{Sec:AppB}  Cavityless system}
We describe the interaction between the particle motion and the free space electromagnetic mode via the Hamiltonian~\cite{magrini_squeezed_2022,militaru_ponderomotive_2022}:
\begin{align}\label{System_Hamiltonian}
	H = \frac{{p_m^2}}{{2m}} + \frac{1}{2}m\omega _m^2x_m^2 + \int {d\omega \,\omega a_\omega ^\dag {a_\omega }}  - G\int {d\omega \,{x_m} (a_\omega ^\dag  + {a_\omega })}
\end{align}
The NP is dispersively coupled to the optical field denoted by the operators ${a}_\omega$ and $a^\dag _\omega $ with optomechanical coupling constant $G$.
In this Hamiltonian, the term $a^\dag_\omega{a}_\omega$ has units of $s^{-1}$ and determines the probability per unit time to count a photon at specified frequency $\omega$ and time $t$.
The NP is also driven by thermal force.
Therefore, we include an effective thermal bath to consider these additional fluctuating forces.
The corresponding Heisenberg-Langevin equations of motion are given by
\begin{subequations}
	\begin{align}
		&\dot x = \omega_m p\,,\label{EQOMa}\\
		&{{\dot p}} = - \omega_m x -\gamma p +  g\int {d\omega ( a^\dag_\omega  + a_\omega )} + \xi (t)\,, \label{EQOMb}\\
		&{\dot a}_\omega =  - i\omega a_\omega  + igx \label{EQOMc}\,.
	\end{align}
\end{subequations}
Here $x = \sqrt {m{\omega _m}} {x_m}$ and $p = {p_m}/\sqrt {m{\omega _m}} $ are the dimensionless position and momentum operators.
Eq.~(\ref{EQOMc}) can be integrated to have a formal solution for the optical mode operator $ a_\omega(t) $. The solutions depend on whether the initial conditions at time $t_-<t$ (the input) or in terms of the final conditions at times $t_+>t$, (the output) are chosen and are given by
\begin{subequations}
	\begin{align}
		a_\omega(t) &= a_\omega (t_ -){e^{i\omega (t_ -  - t)}} + ig\int_{t_ - }^t {ds\,x(s)} e^{i\omega (s - t)}\,,\\
		a_\omega (t) &= a_\omega (t_ + )e^{i\omega ({t_ + } - t)} - ig\int_t^{t_ + } {ds\,x(s)} e^{i\omega (s - t)}\,,
	\end{align}
\end{subequations}
where in each equation the first term describes the free evolution of the optical modes, and the second term arises from their interaction with the NP. By defining mechanical mode vector of operators ${\bf{r}}_m(t)=[p(t),0,x(t),0]^T$, and eliminating the optical mode operators $ a_\omega (t) $ by substituting the formal solution of $ a_\omega (t) $ into Eq.~(\ref{EQOMb}), we have
\begin{align}
	&	{\dot{{\bf{r}}}_m}(t) = {{\bf{A}}_m}{{\bf{r}}_m}(t) + {{\bf{N}}_{{\rm{in}}}}\,,\\
	&	{{\bf{A}}_m} = \left( {\begin{array}{*{20}{c}}
			{ - \gamma }&0&{ - {\omega _m}}&0\\
			0&0&0&0\\
			{{\omega _m}}&0&0&0\\
			0&0&0&0
	\end{array}} \right)\,,\\
	&	{{\bf{N}}_{{\rm{in}}}} = {\left[ {\xi (t) + \sqrt {4\Gamma } {X_{{\rm{in}}}}(t),0,0,0} \right]^T}\,,
\end{align}
where $  X_{\rm in}(t)= (a_{\rm in}(t)+ a_{\rm in}^\dagger(t))/\sqrt{2} $ is the input amplitude quadrature with
\begin{subequations}
	\begin{align}
		a_{\rm in }(t)& = \frac{1}{{\sqrt {2\pi } }}\int {d\omega a_\omega ({t_ - }){e^{i\omega ({t_ - } - t)}}}\,, \\
		a_{\rm out}(t)& = \frac{1}{{\sqrt {2\pi } }}\int {d\omega a_\omega ({t_ + }){e^{i\omega ({t_ + } - t)}}} \,.
	\end{align}
\end{subequations}
Here $\xi (t)$ is the effective thermal force with correlations $\left\langle {\xi (t)\xi (t')} \right\rangle  = 2\gamma \left( {\bar n + 1/2} \right)\delta (t - t')$.
The operators $ X_{\rm out} (t)$ and $ Y_{\rm out} (t)$ describe the transmitted output quadratures and are given by the input-output relation
\begin{subequations}
	\begin{align}
		&{X_{{\rm{out}}}}(t) = {X_{{\rm{in}}}}(t) \,, \\
		&{Y_{{\rm{out}}}}(t) = {Y_{{\rm{in}}}}(t) + \sqrt {4\Gamma } x(t)\,, \label{Eq:Input_Output_2}
	\end{align}
\end{subequations}
where $\Gamma  = 2\pi G^2 x_{zpf}^2$ is the quantum backaction decoherence rate.
The solution for the mechanical mode operators 	${\bf{r}}_m(t)=[p(t),0,x(t),0]^T$ are then given by
\begin{equation}
	{\bf{r}}_m(\tau) = {\bf{M}}(\tau){\bf{r}}_m(0) + \int\limits_0^\tau {ds{\bf{M}}(\tau - s){\bf N}_{{\rm{in}}}(s)}
\end{equation}
where we have defined $	{\bf{M}}(\tau ) = \exp \left[ {  {{\bf{A}}_m}\tau } \right]$ with elements
\begin{align}
	{{\bf{M}}_{11}}(\tau ) &= {e^{ - \gamma \tau /2}}\left( {\cos \Omega \tau - \frac{\gamma }{{2\Omega }}\sin \Omega \tau} \right)\,,\\
	{{\bf{M}}_{33}}(\tau ) &= {e^{ - \gamma \tau /2}}\left( {\cos \Omega \tau + \frac{\gamma }{{2\Omega }}\sin \Omega \tau} \right)\,,\\
	{{\bf{M}}_{13}}(\tau ) &=  - {{\bf{M}}_{31}}(\tau ) = \frac{{{\omega _m}}}{\Omega }{e^{ - \gamma \tau /2}}\sin \Omega \tau\,,\\
	{{\bf{M}}_{22}}(\tau ) &=   {{\bf{M}}_{44}}(\tau ) = 1\,.
\end{align}
With $\Omega  = \frac{1}{2}\sqrt {4\omega _m^2 - {\gamma ^{\rm{2}}}} $.
Input and output light temporal	modes are then defined in the same fashion as Eq.~(17) with a rectangular temporal mode.
\begin{equation}
	[{\cal X}_{\rm in, out}(\tau),{\cal Y}_{\rm in,out}(\tau)] =\frac{1}{\sqrt \tau}\int_{0}^{\tau}[ X_{\rm in, out} (t),Y_{\rm in, out} (t)] dt
\end{equation}
We then write ${\boldsymbol{u}}_q (\tau ) = {[ {{p}(\tau ),{\cal Y}_{\rm out}(\tau ),x(\tau ),{\cal X}_{\rm out} (\tau )} ]^T}$ to calculate the covariance matrix for the cavityless system.

\section{\label{Sec:App2} Elements of covariance matrix}
This section is devoted to the evaluation of the CM elements. In calculating the covariance matrix, we deal with three different types of elements.
\subsection{Cavity-based system}
First, the mechanical elements are calculated according to
\begin{align}
	{{\bf{V}}_{ii'}}(\tau ) =& {{\bf{M}}_{ij}}(\tau ){{\bf{M}}_{i'j'}}(\tau ){{\bf{V}}_{jj'}}(0)+ \int\limits_0^\tau  {ds{{\bf{M}}_{ij}}(\tau  - s){{\bf{M}}_{i'j'}}(\tau  - s){{\bf{N}}_{jk}}{{\bf{N}}_{j'k'}}{\bf{V}}_{kk'}^{{\rm{in}}}}
\end{align}
while cross-correlation between optical and mechanical modes are given by
\begin{align}
	{{\bf{V}}_{ii'}}(\tau ) = &\sqrt {2\kappa } {{\bf{M}}_{ij}}(\tau ){{\bf{V}}_{jj'}}(0)\int\limits_0^\tau  {ds'f(s'){{\bf{M}}_{i'j'}}(s')}- {\bf{V}}_{ki'}^{{\rm{in}}}\int\limits_0^\tau  {ds{{\bf{M}}_{ij}}(\tau  - s){{\bf{N}}_{jk}}} f(s)\nonumber\\
	&+ \sqrt {2\kappa } {\bf{V}}_{kk'}^{{\rm{in}}}\int\limits_0^\tau  {ds{{\bf{M}}_{ij}}(\tau  - s){{\bf{N}}_{jk}}} \int\limits_s^\tau  {ds'f(s'){{\bf{M}}_{i'j'}}(s' - s){{\bf{N}}_{j'k'}}}
\end{align}
and finally, optical correlations have the following form
\begin{align}
	{{\bf{V}}_{ii'}}(\tau ) =& {\bf{V}}_{ii'}^{{\rm{in}}} + 2\kappa {{\bf{V}}_{jj'}}(0)\int\limits_0^\tau  {dsds'f(s)f(s'){{\bf{M}}_{i'j'}}(s){{\bf{M}}_{ij}}(s')} 	- 2\sqrt {2\kappa } {\bf{V}}_{ii'}^{{\rm{in}}}\int\limits_0^\tau  {dsf(s)\int\limits_s^\tau  {ds'f(s'){{\bf{M}}_{i'j'}}(s' - s){{\bf{N}}_{j'k'}}} } \nonumber\\
	&		+ 2\kappa {\bf{V}}_{kk'}^{{\rm{in}}}\int\limits_0^\tau  {ds} \int\limits_s^\tau  {\int\limits_s^\tau  {ds'ds''f(s)f(s'')} } {{\bf{M}}_{i'j'}}(s' - s){{\bf{M}}_{i'j'}}(s'' - s){{\bf{N}}_{jk}}{{\bf{N}}_{j'k'}}
\end{align}

\subsection{Cavityless system}
The mechanical correlations can be calculated as
\begin{equation}
	{{\bf{V}}_{ii'}}(\tau ) = {{\bf{M}}_{ij}}(\tau ){{\bf{M}}_{i'j'}}(\tau ){{\bf{V}}_{jj'}}(0) + \int_0^\tau  {ds{{\bf{M}}_{ij}}(\tau  - s){{\bf{M}}_{i'j'}}(\tau  - s){\bf{V}}_{kk'}^{{\rm{in}}}}
\end{equation}
The cross-correlation between the rectangular temporal optical and mechanical modes are evaluated by
\begin{align}
	{{\bf{V}}_{i2}}(\tau )={{\bf{V}}_{2i}}(\tau )  =& \sqrt {\frac{{4\Gamma }}{\tau }} \left\{  {{{\bf{M}}_{ij}}(\tau ){{\bf{V}}_{jj}}(0)\int\limits_0^\tau  {ds'} {{\bf{M}}_{3j}}(s') + \left[ {2\gamma \left( {\bar n + 1/2} \right) + 2 \Gamma  } \right]\int\limits_0^\tau  {\int\limits_0^\tau  {dsds'{{\bf{M}}_{31}}(s' - s){{\bf{M}}_{i1}}(\tau  - s)} } } \right\}\,,\\
	{{\bf{V}}_{i4}}(\tau ) ={{\bf{V}}_{4i}} (\tau )=& \sqrt {\frac{\Gamma }{\tau }} \int\limits_0^\tau  {ds{{\bf{M}}_{i1}}(\tau  - s)} \,,
\end{align}
and finally, optical correlations are
\begin{align}
	{{\bf{V}}_{22}}(\tau ) =& {\bf{V}}_{22}^{{\rm{in}}} + {{\bf{V}}_{jj'}}(0)\frac{{4\Gamma }}{\tau }\int\limits_0^\tau  {ds} \int\limits_0^\tau  {ds'{{\bf{M}}_{3j}}(s){{\bf{M}}_{3j'}}(s')}  + \frac{{4\Gamma \left[ {2\gamma (\bar n + 1) + 2\Gamma } \right]}}{\tau }\int\limits_0^\tau  {dt} \int\limits_0^{t} {ds\int\limits_s^\tau  {ds'} {{\bf{M}}_{31}}(s' - s){{\bf{M}}_{31}}(t - s)} \\
	{{\bf{V}}_{44}}(\tau ) =& {\bf{V}}_{44}^{{\rm{in}}}\\
	{{\bf{V}}_{42}}(\tau ) =& {{\bf{V}}_{24}}(\tau ) = \frac{{4\Gamma }}{\tau }{\bf{V}}^{{\rm{in}}}_{44}\int\limits_0^\tau  {ds'} \int\limits_0^{s'} {ds{{\bf{M}}_{31}}(s' - s)}
\end{align}

\section{\label{Sec:App3} Derivation of mechanical Wigner function}
The covariance matrix defined in Eq.~(27) can be related to the covariance matrix of the quadratures according to ${{\bf{V}}_q} = \boldsymbol{\Omega} {\bf{V}}{\boldsymbol{\Omega} ^\dag }$. We then relate the Wigner function to the covariance matrix in the quadratures space according to
\begin{equation} \label{Eq:Wigner_quadrature}
	W(\boldsymbol{p},\boldsymbol{q};\rho_{AB}) = \frac{1}{{{\pi ^2}\sqrt {\det( {{\bf{V}}_q})} }}\exp ( - \frac{1}{2}{\boldsymbol{u}}_q^T{\bf{V}}_q^{ - 1}{{\boldsymbol{u}}_q}).
\end{equation}
The density matrix can be written in the basis of coherent states as
\begin{equation} \label{Eq:DM}
	{\rho _{AB}} = \frac{1}{{{\pi ^4}}}\int {{d^2}\boldsymbol{\alpha} {d^2}\boldsymbol{\beta} \left| \boldsymbol{\beta}   \right\rangle } \left\langle \boldsymbol{\beta}  \right|{\rho _{AB}}\left| \boldsymbol{\alpha}  \right\rangle \left\langle \boldsymbol{\alpha}  \right|\,,
\end{equation}
where $\left| \boldsymbol{\alpha}  \right\rangle  = \left| {{\alpha _1},{\alpha _2}} \right\rangle $ and $\left| \boldsymbol{\beta}  \right\rangle  = \left| {{\beta _1},{\beta _2}} \right\rangle $.
Wigner-Weyl transform $W_{\boldsymbol{\alpha}, \boldsymbol{\beta}}(\boldsymbol{p},\boldsymbol{q})$ of the operator $\left| \boldsymbol{\beta}  \right\rangle \left\langle \boldsymbol{\alpha}  \right|$
\begin{align}\label{Eq:Wigner_Weyl}
	W_{\boldsymbol{\alpha}, \boldsymbol{\beta}}(\boldsymbol{p},\boldsymbol{q})	= & 4\exp \left[ { - \frac{{|\boldsymbol{\alpha} {|^2} + |\boldsymbol{\beta} {|^2}}}{2} - {\boldsymbol{\alpha} ^T}{\boldsymbol{\beta} ^*} - {\boldsymbol{p}^T}\boldsymbol{p} - {\boldsymbol{q}^T}\boldsymbol{q}} \right. \nonumber\\
	&	\left. { + \sqrt 2 {\boldsymbol{\alpha} ^T}(\boldsymbol{q} - i\boldsymbol{p}) + \sqrt 2 {\boldsymbol{\beta} ^\dag }(\boldsymbol{q} + i\boldsymbol{p})} \right]\,,
\end{align}
allows us to proceed from the Wigner function to the elements of the density matrix in the coherent state representation by calculating a four-dimensional overlap integral~\cite{dodonov_multidimensional_1994,su_conversion_2019}
\begin{equation}
	\left\langle \boldsymbol{\beta}  \right|{\rho _{AB}}\left| \boldsymbol{\alpha}  \right\rangle=\frac{1}{{4{\pi ^2}}}\int {d\boldsymbol{p}d\boldsymbol{q}{W_{\boldsymbol{\alpha} ,\boldsymbol{\beta} }}(p,q)W(\boldsymbol{p},\boldsymbol{q};\rho_{AB})}
\end{equation}
We use Eq.~(\ref{Eq:Wigner_quadrature}) and perform a Gaussian
integration to obtain
\begin{align}\label{Eq:DM2}
	\rho _{AB} = \frac{{{{\cal P}_0}}}{{{\pi ^4}}}\int {{d^2}\boldsymbol{\alpha} {d^2}\boldsymbol{\beta} \left| \boldsymbol{\beta}  \right\rangle } \left\langle \boldsymbol{\alpha}  \right|{e^{ - \frac{1}{2}{{\left| {\tilde {\boldsymbol{s}}} \right|}^2} + \frac{1}{2}{{\tilde {\boldsymbol{s}}}^T}{\tilde{\bf R}}\tilde {\boldsymbol{s}}}}.
\end{align}
With  $\tilde{\boldsymbol{s}} = (\beta_1^*,\beta_2^*, \alpha_1,\alpha_2)^T$ and $\tilde{\bf R}$ being a $4 \times 4$ symmetric complex matrix written in terms of ${\bf V}$ as
\begin{align}
	&	\tilde{\bf R} =\boldsymbol{\Omega}^T \boldsymbol{\Omega} (2 {\bf V} -  {\bf I}_{4}) (2{\bf V} + {\bf I}_{4} )^{-1},\\
	&	\mathcal{P}_0=\frac{4}{\sqrt{{\rm det}(2{\bf V}+{\bf I}_4)}}.
\end{align}

The unnormalized density matrix $\tilde{\rho}_B(n_2)$ of the mechanical mode can be found by taking the state ${\rho}_{AB}$ and projecting it onto the detected optical mode $|n_2\rangle$. By performing this projection of the state onto $|n_2\rangle$, we obtain $\tilde{\rho}_B(n_2)$ as
\begin{align}
	\tilde \rho_B (n_2)&= \langle n_2 | \rho_{AB}| n_2 \rangle \nonumber\\
	&=
	\frac{\mathcal{P}_0}{\pi^{4}} \int \, \mathrm d^2 {\alpha}_1 \, d^2 \beta_1  \, d^2 \alpha_2  \,\mathrm d^2 \beta_2 ~
	| \beta_1 \rangle \langle \alpha_1 | \, \langle n_2 | \beta_2 \rangle \langle \alpha_2 | n_2 \rangle \, \nonumber\\
	&\qquad\times  \exp\big(-\frac{| \tilde{\boldsymbol{s}} |^2}{2} + \frac{1}{2} \tilde{\boldsymbol{s}}^T \tilde{\bf R} \tilde{\boldsymbol{s}}  \big). \label{eq:DM2}
\end{align}
For a coherent state $\left| {{\alpha _2}} \right\rangle  = {e^{ - |{\alpha _2}{|^2}/2}}\sum\nolimits_{{n_{2 = 0}}}^\infty  {\frac{{\alpha _2^{{n_2}}}}{{\sqrt {{n_2}!} }}\left| {{n_2}} \right\rangle } $ the product $\langle n_2 | \beta_2 \rangle \langle \alpha_2 | n_2 \rangle$ can be calculated as
\begin{eqnarray}\label{Eq:CF}
	\langle n_2 | \beta_2 \rangle \langle \alpha_2 | n_2 \rangle
	= \frac{1}{n_2!} \, e^{-(|\alpha_2|^2 + |\beta_2|^2 )/2}  \big(\alpha_2^* \beta_2 \big)^{n_2},
\end{eqnarray}
To perform the integration, we define vector $\boldsymbol{s}={\bf P} \tilde{\boldsymbol{s}} = (\beta_1^*, \alpha_1, \beta_2^*, \alpha_2)^T= (\boldsymbol{s}_m, \boldsymbol{s}_c)^T$, where $\boldsymbol{s}_m$ and $\boldsymbol{s}_c$ are vectors corresponding to the mechanical mode and optical modes, respectively and ${\bf P}$ is the permutation matrix. Correspondingly, we define a new symmetric matrix ${\bf R}$  as
\begin{equation} \label{eq:R}
	{\bf{R}} = {\bf{P}}\tilde {\bf{R}}{{\bf{P}}^T} = \left[ {\begin{array}{*{20}{c}}
			{{{\bf{R}}_{mm}}}&{{{\bf{R}}_{mc}}}\\
			{{{\bf{R}}_{cm}}}&{{{\bf{R}}_{cc}}}
	\end{array}} \right],
\end{equation}
where ${\bf R}_{mm}$ and ${\bf R}_{cc}$ are $2 \times 2$ symmetric matrices corresponding to the mechanical and optical modes respectively and ${\bf R}_{mc}$ is a $2 \times 2$ matrix that connects the two modes. Since ${\bf R}$ is symmetric, ${\bf R}_{cm} = {\bf R}_{mc}^T$.
\begin{align}
	&| \tilde{\boldsymbol{s}} |^2 = | \boldsymbol{s}_m |^2 + | \boldsymbol{s}_c |^2,
	\nonumber\\
	&\tilde{\boldsymbol{s}}^T \tilde{\bf R} \tilde{\boldsymbol{s}} =
	\boldsymbol{s}_m^T {\bf R}_{mm} \boldsymbol{s}_m + \boldsymbol{s}_c^T {\bf R}_{cc} \boldsymbol{s}_c
	+ 2 \, \boldsymbol{s}_m^T {\bf R}_{mc} \boldsymbol{s}_c\,, \label{Eq:RS}
\end{align}
Inserting Eqs.~(\ref{Eq:CF}) and (\ref{Eq:RS}) into Eq.~(\ref{eq:DM2}), the unnormalized density matrix $\tilde \rho_B(n)$ can be written as
\begin{eqnarray}\label{eq:DM-3}
	\tilde \rho_B(n_2) = \frac{1}{\pi^{2}} \int  d^2 {\alpha}_1   d^2 \beta_1  | \beta_1 \rangle \langle \alpha_1 | F(\alpha_1, \beta_1),
\end{eqnarray}
where
\begin{align}\label{eq:Ffunction}
	F({\alpha _1},{\beta _1}) &= \frac{{{{\cal P}_0}{\mkern 1mu} }}{{{\pi ^2}{n_2}!}}{e^{ - \frac{1}{2}|{{\bf{s}}_m}{|^2} + \frac{1}{2}{\bf{s}}_m^T{{\bf{R}}_{mm}}{{\bf{s}}_m}}}\int {{d^2}{\alpha _2}{\mkern 1mu} {d^2}{\beta _2}\;{{(\alpha _2^*{\beta _2})}^{{n_2}}}{e^{ - |{{\bf{s}}_c}{|^2} + \frac{1}{2}{\bf{s}}_c^T{{\bf{R}}_{cc}}{{\bf{s}}_c} + {\bf{s}}_c^T{{\bf{R}}_{cm}}{{\bf{s}}_m}}}} \nonumber\\
	&	= \frac{{{{\cal P}_0}}}{{{n_2}!}}{\mkern 1mu} {e^{ - \frac{1}{2}|{{\bf{s}}_m}{|^2} + \frac{1}{2}{\bf{s}}_m^T{{\bf{R}}_{mm}}{{\bf{s}}_m}}}{(\frac{{{\partial ^2}}}{{\partial {\alpha _2}\partial \beta _2^*}})^{{n_2}}}{e^{ - |{{\bf{s}}_c}{|^2} + \frac{1}{2}{\bf{s}}_c^T{{\bf{R}}_{cc}}{{\bf{s}}_c} + {\bf{s}}_c^T{{\bf{R}}_{cm}}{{\bf{s}}_m}}}{|_{{{\bf{s}}_c} = {\bf{0}}}},
\end{align}
We next calculate the unnormalized characteristic function $\chi_B^{n_2}(\beta)$
\begin{align}
	\chi_B^{n_2}(\beta)
	&=e^{-\frac{|\beta|^2}{2}} \text{Tr}\big( e^{ -\beta^*  a} \tilde \rho_B(n_2) e^{\beta { a}^{\dag}}  \big) \nonumber\\
	&=\frac{e^{-\frac{|\beta|^2}{2}}}{\pi^{2}}  \int \mathrm d^2 {\alpha}_1 \mathrm d^2 \beta_1
	e^{ \beta \alpha_1^* - \beta^* \beta_1 } \langle \alpha_1  | \beta_1 \rangle F(\alpha_1, \beta_1),
	\label{Eq:characteristic}
\end{align}
In the coherent state basis, the Wigner function for the single-mode state is defined as
\begin{equation}\label{eq:Wigner1}
	W_B^{n_2}(\alpha,\rho_B) = \frac{1}{{{\pi ^2}}}\int  {d^2}\beta {e^{-\beta {\alpha ^*} + \alpha {\beta ^*}}}\chi^{n_2}_B (\beta ),
\end{equation}
Substituting  $\chi_B^{n_2}(\beta)$ from Eq.~(\ref{Eq:characteristic}) into the definition of Wigner function Eq.~(\ref{eq:Wigner1}) we find the unnormalized Wigner function as
\begin{align}
	W_B^{n_2}(\alpha) =&\frac{1}{\pi^{4}}   \int  d^2 {\alpha}_1   d^2 \beta_1  d^2 \beta  \langle \alpha_1  | \beta_1 \rangle F(\alpha_1, \beta_1)
	~ e^{-|\beta|^2/2} e^{ - \beta^* (\beta_1-\alpha) +  \beta (\alpha_1^*-\alpha^*)}  \nonumber\\
	=&
	\frac{2}{\pi^{3}}  e^{-2|\alpha|^2} \int \mathrm d^2 {\alpha}_1  \mathrm d^2 \beta_1 ~ F(\alpha_1, \beta_1) \,
	e^{-\frac{|\alpha_1|^2}{2} - \frac{|\beta_1|^2}{2} - \alpha_1^* \beta_1 + 2 \, (\alpha \alpha_1^*+\alpha^* \beta_1) },
	\label{eq:Wigner3}
\end{align}
where in the last equality we have performed the integration over $\beta$ and used the relation
$\langle \alpha_1  | \beta_1 \rangle = e^{-|\alpha_1|^2/2 - |\beta_1|^2/2 + \alpha_1^* \beta_1}$.
By substituting the function $F(\alpha_1, \beta_1)$ of Eq.~(\ref{eq:Ffunction}) into Eq.~(\ref{eq:Wigner3}), interchanging the order of partial derivatives
and integration, and then performing the integration over $\alpha_1$ and $\beta_1$, we arrive at the final expression of Eq.~(29) in the main text for the unnormalized Wigner function of the mechanical mode.

\section{Non-Classicality of heralded mechanical states} 
\label{sec:non_classicality_of_heralded_mechanical_states}

\begin{figure}
	\centering
	\includegraphics[scale=1]{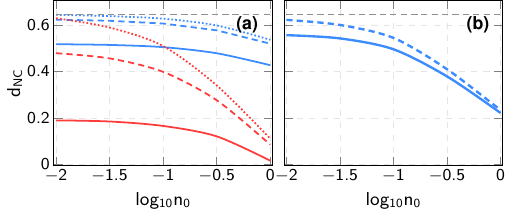}
	\caption{
		Non-classical depth as a function of initial thermal occupation of the mechanical oscillator $n_0$.
		(a) Phonon-added (blue lines) and subtracted (red lines) and (b) cavityless system.
		Each curve corresponds to a different mechanical state marked in Fig.~2 and Fig.~3.
		The upper bound corresponds to a perfect single phonon state shown by the dashed gray line.
	}
	\label{FigS3}
\end{figure}

To complete our evaluation of the heralded mechanical states, we use the following two algebraic inequalities to certify nonclassicality from a few Fock-state probabilities that are well-suited to experimental implementations~\cite{innocenti_nonclassicality_2022}:
\begin{align}
	&	Q_1^2 > 2{Q_0}{Q_2}\,, \label{NC1}\\
	&{Q_0} + \frac{{Q_1^2}}{{2{Q_2}}}\left[ {\exp \left( {\frac{{2{Q_2}}}{{Q_1}}} \right) - 1} \right] > 1 \,. \label{NC2}
\end{align}
We have checked that only criterion inequality~(\ref{NC1}) is capable of detecting the non-classicality of the generated state, while inequality~(\ref{NC2}) is not applicable.

Similarly to QNG depth, we define the quantum non-classical depth, $d_{NC}$, quantifying the amount of heating required to reach the non-classical boundary given by Eqs.~(\ref{NC1}) and (\ref{NC2}).
It is worth noting that for an ideal single phonon state $\rho_B=\left| 1 \right\rangle \left\langle 1 \right|$, the non-classical depth is found to be $d_{\rm NC}=0.646$, represented by a dashed gray line.
Fig.~\ref{FigS3} shows the non-classical depth versus the initial mechanical occupation for various mechanical states marked in Figs.~2 and 3. This value is approximately twice the magnitude of the corresponding non-Gaussian depth.


%

\end{document}